\input harvmac\skip0=\baselineskip
\divide\skip0 by 2

\noblackbox
\def\cp{{\cal P}}
\def\cc{{\cal C}}
\def\ct{{\cal T}}
\def\cU{{\cal U}}
\def\ipm{$ { \cal I}^\pm$}
\def\opm{{\cal O}_\pm }
\def\om{{\cal O}_-}
\def\op{{\cal O}_+}
\def\t{\tau}
\def\s{\Omega}
\def\half{{1\over2}}
\def\a{\alpha}

\def\ep{\epsilon}

\def\G{\Gamma}
\def\p{\partial}
\def\bw{{\bar w}}

\def\ds{dS$_3$}
\def\hi{$\hat { \cal I}^-$}
\def\im{$ { \cal I}^-$}
\def\ip{$ { \cal I}^+$}
\def\ri{{\rm in}}
\def\ro{{\rm out}}
\def\msurr{\mathsurround=0pt}
\def\overleftrightarrow#1{\vbox{\msurr\ialign{##\crcr
        $\leftrightarrow$\crcr\noalign{\kern-1pt\nointerlineskip}
        $\hfil\displaystyle{#1}\hfil$\crcr}}}
\def\csch{\mathop{\rm csch}}
\def\hi{$\hat { \cal I}^-$}
\def\ev{|E\rangle}
\def\av{|\a\rangle}
\def\iv{|\ri \rangle}
\def\ov{|\ro \rangle}
\def\half{{1\over2}}

\lref\nm{M. Spradlin and A. Volovich, to appear.}
\lref\StromingerGP{
A.~Strominger,
``Inflation and the dS/CFT correspondence,''
arXiv:hep-th/0110087.
}
\lref\CacciatoriUN{
S.~Cacciatori and D.~Klemm,
``The asymptotic dynamics of de Sitter gravity in three dimensions,''
arXiv:hep-th/0110031.
}
\lref\MyungAB{
Y.~S.~Myung,
``Entropy of the three dimensional Schwarzschild-de Sitter black hole,''
arXiv:hep-th/0110123.
}
\lref\CarneirodaCunhaNW{
B.~G.~Carneiro da Cunha,
``Three-dimensional de Sitter gravity and the correspondence,''
arXiv:hep-th/0110169.
}
\lref\CaiJD{
R.~G.~Cai, Y.~S.~Myung and Y.~Z.~Zhang,
``Check of the mass bound conjecture in de Sitter space,''
arXiv:hep-th/0110234.
}
\lref\OgushiIJ{
S.~Ogushi,
``Holographic entropy on the brane in de Sitter Schwarzschild space,''
arXiv:hep-th/0111008.
}

\lref\tsamis{
N.~C.~Tsamis and R.~P.~Woodard,
``The quantum gravitational back-reaction on inflation,''
Annals Phys.\  {\bf 253}, 1 (1997)
[hep-ph/9602316].}

\lref\park{
M.~I.~Park,
``Statistical entropy of three-dimensional Kerr-de Sitter space,''
Phys.\ Lett.\ B {\bf 440}, 275 (1998)
[hep-th/9806119].
}
\lref\li{
M.~Li,
``Matrix model for de Sitter,''
arXiv:hep-th/0106184.
}
\lref\nojiri{
S.~Nojiri and S.~D.~Odintsov,
``Conformal anomaly from dS/CFT correspondence,''
Phys.\ Lett.\ B {\bf 519}, 145 (2001)
[arXiv:hep-th/0106191].
}
\lref\noo{
S.~Nojiri, S.~D.~Odintsov and S.~Ogushi,
``Cosmological and black hole brane world universes in higher derivative  gravity,''
arXiv:hep-th/0108172.
}
\lref\figari{
R.~Figari, R.~Hoegh-Krohn and C.~R.~Nappi,
``Interacting Relativistic Boson Fields In The De Sitter
Universe With Two Space-Time Dimensions,''
Commun.\ Math.\ Phys.\  {\bf 44}, 265 (1975).
}
\lref\mazur{
P.~O.~Mazur and E.~Mottola,
``Weyl cohomology and the effective action for conformal anomalies,''
Phys.\ Rev.\ D {\bf 64}, 104022 (2001)
[arXiv:hep-th/0106151].
}
\lref\pns{
K.~Pilch, P.~van Nieuwenhuizen and M.~F.~Sohnius,
``De Sitter Superalgebras And Supergravity,''
Commun.\ Math.\ Phys.\  {\bf 98}, 105 (1985).
}
\lref\eva{
E.~Silverstein,
``(A)dS backgrounds from asymmetric orientifolds,''
hep-th/0106209.
}
\lref\chamblin{
A.~Chamblin and N.~D.~Lambert,
``Zero-branes, quantum mechanics and the cosmological constant,''
hep-th/0107031.
}
\lref\gao{
Y.~Gao,
``Symmetries, matrices, and de Sitter gravity,''
hep-th/0107067.
}
\lref\cardy{
J.~L.~Cardy,
``Operator Content Of Two-Dimensional Conformally Invariant Theories,''
Nucl.\ Phys.\ B {\bf 270}, 186 (1986).
}

\lref\bros{
J.~Bros, H.~Epstein and U.~Moschella,
``The asymptotic symmetry of de Sitter spacetime,''
hep-th/0107091.
}
\lref\nojiriod{
S.~Nojiri and S.~D.~Odintsov,
``Quantum cosmology, inflationary brane-world creation and dS/CFT
correspondence,''
hep-th/0107134.
}
\lref\halyo{
E.~Halyo,
``De Sitter entropy and strings,''
hep-th/0107169.
}
\lref\tolley{
A.~J.~Tolley and N.~Turok,
``Quantization of the massless minimally coupled 
scalar field and the dS/CFT correspondence,''
hep-th/0108119.
}
\lref\sit{
T.~Shiromizu, D.~Ida and T.~Torii,
``Gravitational energy, dS/CFT correspondence and cosmic no-hair,''
hep-th/0109057.
}
\lref\kallosh{
R.~Kallosh,
``N = 2 supersymmetry and de Sitter space,''
hep-th/0109168.
}
\lref\hull{
C.~M.~Hull,
``De Sitter Space in Supergravity and M Theory,''
hep-th/0109213.
}
\lref\abbott{
L.~F.~Abbott and S.~Deser,
``Stability Of Gravity With A Cosmological Constant,''
Nucl.\ Phys.\ B {\bf 195}, 76 (1982).
}
\lref\cht{
N.~A.~Chernikov and E.~A.~Tagirov,
``Quantum Theory Of Scalar Fields In De Sitter Space-Time,''
Annales Poincare Phys.\ Theor.\ A {\bf 9} (1968) 109.
}
\lref\tagirov{
E.~A.~Tagirov,
``Consequences Of Field Quantization In De Sitter Type Cosmological Models,''
Annals Phys.\  {\bf 76}, 561 (1973).
}
\lref\rumpf{
H.~Rumpf and H.~K.~Urbantke,
``Covariant `In-Out' Formalism For Creation By External Fields,''
Annals Phys.\  {\bf 114}, 332 (1978).
}
\lref\najmi{
A.~H.~Najmi and A.~C.~Ottewill,
``Quantum States And The Hadamard
Form. I. Energy Minimization For Scalar Fields,''
Phys.\ Rev.\ D {\bf 30}, 1733 (1984).
}
\lref\ford{
L.~H.~Ford,
``Quantum Instability Of De Sitter Space-Time,''
Phys.\ Rev.\ D {\bf 31}, 710 (1985).
}
\lref\allenfol{
B.~Allen and A.~Folacci,
``The Massless Minimally Coupled Scalar Field In De Sitter Space,''
Phys.\ Rev.\ D {\bf 35}, 3771 (1987).
}
\lref\cai{
R.~G.~Cai,
``Cardy-Verlinde Formula and Asymptotically de Sitter Spaces,''
arXiv:hep-th/0111093.
}

\lref\unruh{
W.~G.~Unruh,
``Notes On Black Hole Evaporation,''
Phys.\ Rev.\ D {\bf 14}, 870 (1976).
}
\lref\ulf{
U.~H.~Danielsson,
``A black hole hologram in de Sitter space,''
arXiv:hep-th/0110265.
}
\lref\idet{
I.~Sachs and S.~N.~Solodukhin,
``Horizon holography,''
arXiv:hep-th/0107173.
}
\lref\bdm{
V.~Balasubramanian, J.~de Boer and D.~Minic,
``Mass, entropy and holography in asymptotically de Sitter spaces,''
arXiv:hep-th/0110108.
}
\lref\klemm{
D.~Klemm,
``Some aspects of the de Sitter/CFT correspondence,''
[hep-th/0106247].
}

\lref\petkou{
A.~C.~Petkou and G.~Siopsis,
``dS/CFT correspondence on a brane,''
arXiv:hep-th/0111085.
}
\lref\dscft{A.~Strominger,
``The dS/CFT Correspondence,''
hep-th/0106113.}
\lref\coss{		
O.~Coussaert, M.~Henneaux and P.~van Driel,
``The Asymptotic dynamics of three-dimensional
Einstein gravity with a negative cosmological constant,''
Class.\ Quant.\ Grav.\  {\bf 12}, 2961 (1995)
[gr-qc/9506019].}
\lref\deja{
S.~Deser and R.~Jackiw,
``Three-Dimensional Cosmological Gravity:
Dynamics Of Constant Curvature,''
Annals Phys.\  {\bf 153}, 405 (1984).}
\lref\hms{
S.~Hawking, J.~Maldacena and A.~Strominger,
``DeSitter entropy, quantum entanglement and AdS/CFT,''
JHEP {\bf 0105}, 001 (2001)
[hep-th/0002145].}
\lref\bousso{
R.~Bousso,
``Bekenstein bounds in de Sitter and flat space,''
JHEP {\bf 0104}, 035 (2001)
[hep-th/0012052].}
\lref\hulla{
C.~M.~Hull,
``Timelike T-duality, de Sitter space, large N gauge theories and
topological field theory,'' JHEP {\bf 9807}, 021 (1998)
[hep-th/9806146].}
\lref\hullb{
C.~M.~Hull and R.~R.~Khuri,
``Worldvolume theories, holography, duality and time,''
Nucl.\ Phys.\ B {\bf 575}, 231 (2000)
[hep-th/9911082].}
\lref\juan{J.~Maldacena,
``The large N limit of superconformal field theories and supergravity,''
Adv.\ Theor.\ Math.\ Phys.\  {\bf 2}, 231 (1998)
[hep-th/9711200].}
\lref\burges{
C.~J.~Burges,
``The De Sitter Vacuum,''
Nucl.\ Phys.\ B {\bf 247}, 533 (1984).}
\lref\jmas{
J.~Maldacena and A.~Strominger,
``Statistical entropy of de Sitter space,''
JHEP {\bf 9802}, 014 (1998)
[gr-qc/9801096].}
\lref\asbh{A.~Strominger,
``Black hole entropy from near-horizon microstates,''
JHEP {\bf 9802}, 009 (1998)
[hep-th/9712251].}
\lref\by{
J.~D.~Brown and J.~W.~York,
``Quasilocal energy and conserved charges derived from the
gravitational action,'' Phys.\ Rev.\ D {\bf 47}, 1407 (1993).}
\lref\bk{V.~Balasubramanian and P.~Kraus,
``A stress tensor for anti-de Sitter gravity,''
Commun.\ Math.\ Phys.\  {\bf 208}, 413 (1999)
[hep-th/9902121].}
\lref\bh{
J.~D.~Brown and M.~Henneaux,
``Central Charges In The Canonical Realization Of Asymptotic
Symmetries:  An Example From Three-Dimensional Gravity,''
Commun.\ Math.\ Phys.\  {\bf 104}, 207 (1986).}
\lref\gkp{
S.~S.~Gubser, I.~R.~Klebanov and A.~M.~Polyakov,
``Gauge theory correlators from non-critical string theory,''
Phys.\ Lett.\ B {\bf 428}, 105 (1998)
[arXiv:hep-th/9802109].
}
 \lref\witt{E.~Witten,
``Anti-de Sitter space and holography,''
Adv.\ Theor.\ Math.\ Phys.\  {\bf 2}, 253 (1998)
[hep-th/9802150].}
\lref\hs{M.~Henningson and K.~Skenderis,
``The holographic Weyl anomaly,''
JHEP {\bf 9807}, 023 (1998)
[hep-th/9806087].}
\lref\ad{L.~F.~Abbott and S.~Deser,
``Stability Of Gravity With A Cosmological Constant,''
Nucl.\ Phys.\ B {\bf 195}, 76 (1982).}
\lref\mottolb{I. Antoniadis, P. O. Mazur and E.~Mottola 
``Comment on *Nongaussian isociirvature perturbations from inflation*'',
astro-ph/9705200.}
\lref\mottola{E.~Mottola,
``Particle Creation In De Sitter Space,''
Phys.\ Rev.\ D {\bf 31}, 754 (1985)}
\lref\allen{B.~Allen,
``Vacuum States In De Sitter Space,''
Phys.\ Rev.\ D {\bf 32}, 3136 (1985)}
\lref\birrel{N. D. Birrel and P. C. W.
Davies, {\it Quantum Fields in Curved Space}, Cambridge University
Press, Cambridge (1982).   }
\lref\abramowitz{M. Abramowitz and I. Stegun,
{\it Handbook of Mathematical Functions}, Dover (1972).}
\lref\gradsh{I. S. Gradshteyn and I. M. Ryzhik,
{\it Table of Integrals, Series and Products},
Academic Press (1994).}
\lref\sei{N.~Seiberg,
``Notes On Quantum Liouville Theory And Quantum Gravity,''
Prog.\ Theor.\ Phys.\ Suppl.\  {\bf 102}, 319 (1990).}
\lref\giddings{S.~B.~Giddings,
``The boundary S-matrix and the AdS to CFT dictionary,''
Phys.\ Rev.\ Lett.\  {\bf 83}, 2707 (1999)
[hep-th/9903048].}
\lref\ascv{
A.~Strominger and C.~Vafa,
``Microscopic Origin of the Bekenstein-Hawking Entropy,''
Phys.\ Lett.\ B {\bf 379}, 99 (1996)
[hep-th/9601029].}
\lref\bek{
J.~D.~Bekenstein,
``Black Holes And Entropy,''
Phys.\ Rev.\ D {\bf 7}, 2333 (1973).}
\lref\banksb{
T.~Banks,
``Cosmological breaking of supersymmetry or little Lambda goes back to
the future. II,''
hep-th/0007146.}
\lref\bousso{
R.~Bousso,
``Bekenstein bounds in de Sitter and flat space,''
JHEP {\bf 0104}, 035 (2001)
[hep-th/0012052].}
\lref\boussob{
R.~Bousso,
``Positive vacuum energy and the N-bound,''
JHEP {\bf 0011}, 038 (2000)
[hep-th/0010252].}
\lref\boussoc{
R.~Bousso,
``Holography in general space-times,''
JHEP {\bf 9906}, 028 (1999)
[hep-th/9906022].}
\lref\hawk{S.~W.~Hawking,
``Particle Creation By Black Holes,''
Commun.\ Math.\ Phys.\  {\bf 43}, 199 (1975).}
\lref\vijay{
V.~Balasubramanian, P.~Horava and D.~Minic,
``Deconstructing de Sitter,''
JHEP {\bf 0105}, 043 (2001)
[hep-th/0103171].}
\lref\gibbons{
G.~W.~Gibbons and S.~W.~Hawking,
``Cosmological Event Horizons, Thermodynamics, And Particle Creation,''
Phys.\ Rev.\ D {\bf 15}, 2738 (1977).}
\lref\nastya{
A.~Volovich,
``Discreteness in deSitter space and quantization of Kahler manifolds,''
hep-th/0101176.}
\lref\britto{
R.~Britto-Pacumio, A.~Strominger and A.~Volovich,
``Holography for coset spaces,''
JHEP {\bf 9911}, 013 (1999)
[hep-th/9905211].}
\lref\at{
A.~Achucarro and P.~K.~Townsend,
``A Chern-Simons Action For Three-
Dimensional Anti-De Sitter Supergravity Theories,''
Phys.\ Lett.\ B {\bf 180}, 89 (1986).}
\lref\wu{
F.~Lin and Y.~Wu,
``Near-horizon Virasoro symmetry and the entropy of de
Sitter space in  any dimension,''
Phys.\ Lett.\ B {\bf 453}, 222 (1999)
[hep-th/9901147].}
\lref\kim{
W.~T.~Kim,
``Entropy of 2+1 dimensional de Sitter space in terms of brick wall
method,'' Phys.\ Rev.\ D {\bf 59}, 047503 (1999)
[hep-th/9810169].}
\lref\banados{
M.~Banados, T.~Brotz and M.~E.~Ortiz,
``Quantum three-dimensional de Sitter space,''
Phys.\ Rev.\ D {\bf 59}, 046002 (1999)
[hep-th/9807216].}
\lref\gaowald{
S.~Gao and R.~M.~Wald,
``Theorems on gravitational time delay and related issues,''
Class.\ Quant.\ Grav.\  {\bf 17}, 4999 (2000)
[gr-qc/0007021].}
\lref\cpec{See e.g. S.~Perlmutter,
``Supernovae, dark energy, and the accelerating universe: The status
of the cosmological parameters,''
in {\it Proc. of the 19th Intl. Symp. on Photon and Lepton
Interactions at High Energy LP99 } ed. J.A. Jaros and M.E. Peskin,
Int.\ J.\ Mod.\ Phys.\ A {\bf 15S1}, 715 (2000).}
\lref\boussob{
R.~Bousso,
``Positive vacuum energy and the N-bound,''
JHEP {\bf 0011}, 038 (2000)
[hep-th/0010252].}
\lref\witcs{
E.~Witten,
``(2+1)-Dimensional Gravity As An Exactly Soluble System,''
Nucl.\ Phys.\ B {\bf 311}, 46 (1988).}
\lref\witb{
E.~Witten,
``Quantization of Chern Simon Theory with Complex Gauge Group,''
Commun.\ Math.\ Phys.\  {\bf 137}, 29 (1991).}
\lref\banks{
T.~Banks and W.~Fischler,
``M-theory observables for cosmological space-times,''
hep-th/0102077.}
\lref\witc{
E.~Witten,
``Quantum gravity in de Sitter space,''
arXiv:hep-th/0106109.
}
\lref\witst{ E.~Witten, ``Quantum gravity
in de Sitter space'' Strings 2001 online proceedings
http://theory.theory.tifr.res.in/strings/Proceedings }
\lref\fiopre{
T.~M.~Fiola, J.~Preskill, A.~Strominger and S.~P.~Trivedi,
``Black Hole Thermodynamics and Information Loss in Two Dimensions,''
Phys.\ Rev.\ D {\bf 50}, 3987 (1994)
[hep-th/9403137].}
\lref\bouhaw{
R.~Bousso and S.~W.~Hawking,
``Pair Creation of Black Holes During Inflation,''
Phys.\ Rev.\ D {\bf 54}, 6312 (1996)
[gr-qc/9606052].}
\lref\lectures{
M.~Spradlin, A.~Strominger and A.~Volovich,
``Les Houches lectures on de Sitter space,''
hep-th/0110007.
}
\lref\mcn{
B.~McInnes,
``Exploring the similarities of the dS/CFT and AdS/CFT correspondences,''
arXiv:hep-th/0110062.
}
\lref\birdav{
N. D. Birrell and P. C. W. Davies,
{\it Quantum Fields in Curved Space,}
Cambridge University Press, Cambridge (1982).
}

\Title{\vbox{\baselineskip12pt\hbox{hep-th/0112218}
\hbox{}}}{Conformal Vacua and Entropy in de Sitter Space}

\centerline{Raphael Bousso\footnote{$^*$}{Institute for Theoretical
Physics, University of California, Santa Barbara, CA 93106}, Alexander 
Maloney\footnote{$^\dagger$}{Department of Physics, Harvard University,
Cambridge, MA 02138} and Andrew Strominger$^\dagger$}

\vskip .3in \centerline{\bf Abstract}
{
The dS/CFT correspondence is illuminated through an 
analysis of massive scalar field theory in $d$-dimensional 
de Sitter space.  
We consider a one-parameter family of dS-invariant vacua related by
Bogolyubov transformations and compute the corresponding Green
functions.  It is shown that none of these Green functions correspond
to the one obtained by analytic continuation from AdS.  Among this
family of vacua are in (out) vacua which have no incoming (outgoing)
particles on \im\ (\ip). Surprisingly, it is shown that in odd
spacetime dimensions the in and out vacua are the same, implying the
absence of particle production for this state.  The correlators of the
boundary CFT, as defined by the dS/CFT correspondence, are shown to
depend on the choice of vacuum state---the correlators with all
points on \im\ vanish in the in vacuum.  For \ds\ we argue that this
bulk vacuum dependence of the correlators is dual to a deformation of
the boundary CFT$_2$ by a specific marginal operator. It is also shown that 
Witten's non-standard de Sitter 
inner product (slightly modified) reduces to the standard inner 
product of the boundary field theory.  Next we
consider a scalar field in the Kerr-\ds\ Euclidean vacuum.  A density
matrix is constructed by tracing out over modes which are causally
inaccessible to a single geodesic observer.  This is shown to be a
thermal state at the Kerr-\ds\ temperature and angular potential.  It
is further shown that, assuming Cardy's formula, the microscopic
entropy of such a thermal state in the boundary CFT precisely 
equals the Bekenstein-Hawking value of one quarter the area of the
Kerr-\ds\ horizon. }

\smallskip
\Date{}
\listtoc
\writetoc

\vfill \eject


\newsec{Introduction and Summary}
Recently, following earlier work \refs{\mottolb\hulla\jmas\wu\kim\bousso
\vijay\banados
\boussoc\banksb\boussob{--}\witst}, a proposal has been made relating
quantum gravity in de Sitter space to conformal field theory on the
spacelike boundary of de Sitter space \dscft. The proposal was
motivated by an analysis of the asymptotic symmetry group of de Sitter
space together with an appropriately crafted analogy to the AdS/CFT
correspondence \refs{\juan,\gkp,\witt}. 
Other relevant discussions of quantum gravity in de Sitter space and
dS/CFT appear in
\refs{\witc\eva \nojiri\li\chamblin\gao\bros\nojiriod\tolley 
\nastya\sit\kallosh\tsamis\petkou\mazur\noo\CacciatoriUN
\MyungAB\CarneirodaCunhaNW\CaiJD\OgushiIJ\StromingerGP{--}\hull
}.

Unlike the AdS/CFT case, there has been no derivation of the
proposed dS/CFT correspondence from string theory. Hopefully, a
stringy construction of de Sitter space will be forthcoming.
Meanwhile, much has been learned about AdS/CFT by analyzing
solutions of the field equations and studying the propagation and
interactions of fields, without directly using string theory. In
this paper we pursue a parallel approach to dS/CFT, analyzing in
some detail massive scalar field theory in de Sitter space. A
number of surprising and interesting features emerge. Since this
paper contains some rather detailed calculations, for the benefit
of the reader we include a summary in this introduction.

We begin in section 2 with a discussion of dS-invariant Green
functions for a massive scalar, reviewing and generalizing to $d$
dimensions the discussion of \refs{\mottola,\allen}.  We first
describe the Green function obtained by analytic continuation from the
Euclidean sphere.  This is the so-called Euclidean Green function, and
it is the two-point function of the scalar field in the Euclidean
vacuum. We then construct a family of dS-invariant vacua labeled by a
complex parameter $\alpha$ and compute the Green functions in these
$\alpha$-vacua, which have several peculiarities.  Singularities occur
at antipodal points which are however,
unobservable since antipodal points are
always separated by a horizon.  Moreover, these singularities do not
affect the scalar commutator, which is independent of $\alpha$. We
also see that the coincident point singularity has two terms, with
opposite-signed $i\epsilon$ prescriptions. Hence all of these
$\alpha$-vacua except for the Euclidean vacuum differ from the usual
Minkowski vacuum at arbitrarily short distances. We also compute the
response of an Unruh detector and find that it is thermal only in the
Euclidean vacuum. The dual CFT interpretation of the $\alpha$-vacua is
deferred to section 4.

In relating the AdS/CFT and dS/CFT correspondences, it is natural to
consider the particular Green function obtained by `double' analytic
continuation from AdS to dS via the hyperbolic plane. We show that the
Green function so obtained, while dS-invariant, does {\it not\/}
correspond to the Green function in any known dS-invariant
vacuum.\foot{We benefitted greatly from discussions with M. Spradlin
and A. Volovich on this point. There is in fact a
four-complex-parameter family of dS-invariant Wightman functions,
characterized by the (complex) strengths of the coincident and
antipodal poles, as well as the two possible $i\epsilon$ prescriptions
at each pole.  Only a one-complex-parameter family of these is known
to be realizable as two-point vacuum expectation values. Analytic
continuation from AdS gives a result which is not realized within this
family.}  This result underscores the non-triviality of extrapolating
from AdS/CFT to dS/CFT.

In section 3 we consider scalar field theory in spherical
coordinates \eqn\dsmet{{ds^2\over \ell^2  }=-d\t^2 + \cosh^2\t\,
d\Omega_{d-1}^2,} again generalizing \refs{\mottola, \allen} to $d$
dimensions. A salient feature of these coordinates is that they
cover all of de Sitter space and hence are suitable for studying
global properties.   The solutions of the massive scalar wave
equation are found for arbitrary angular momentum. We then give an
explicit construction in terms of these modes of the Bogolyubov
transformations relating all the $\alpha$-vacua. Special `in' and
`out' vacua are found, which are distinct from the Euclidean
vacuum. The in vacuum has no incoming particles on \im, while the
out vacuum has no outgoing particles on \ip. The Bogulubov
transformation between them is computed. Surprisingly, it is found
to be trivial in odd-dimensions. This means that for the in vacuum
of odd-dimensional de Sitter space there is no particle
production. This result did not appear in previous analyses, which
largely considered the four-dimensional case.

In section 4 we specialize to \ds\ and consider the dual CFT$_2$
interpretation of these results, along the lines proposed in
\dscft. We first compute the boundary behavior of the massive
scalar Green function as a function of the vacuum
parameter $\alpha$. This behavior is fixed by conformal invariance
up to overall constants which are $\alpha$-dependent. The boundary
correlators have an especially simple form in the in vacuum. For
both points on \im\ (or both on \ip) they vanish!\foot{Except for
a contact term which is computed.} This is related to the fact
that on \im\ the spatial kinetic terms vanish and
the theory becomes ultralocal.  For one point on \im\ and one on
\ip\ they do not vanish. The simplicity of this behavior suggests
that the in vacuum, despite the unphysical singularities, 
may play an important role in understanding the
dS/CFT correspondence.

One way of generating a family of correlators in a CFT is by deforming
the theory by a marginal operator.  In \dscft\ it was argued that a
scalar field of mass $m$ is dual to a pair of CFT operators ${\cal
O}_\pm$ with conformal weights $1\pm \sqrt{1-m^2\ell^2}$. The
composite operator ${\cal O}_+{\cal O}_-$ always has dimension 2 
dor any $m$, exactly
what is required for a marginal deformation. We show explicitly for
real $\alpha$ that this composite operator deforms the correlators in
the same way as shifting $\alpha$.

In section 5 we consider the definition of the adjoint in the Hilbert
space of the scalar field. In standard treatments of 2D Euclidean
conformal field theory, the adjoint of an operator involves a
(non-local) reflection about the unit circle. This prescription
becomes the usual local adjoint when mapped to the cylinder.  The
``naive'' adjoint for a bulk scalar field induces an adjoint in the
Euclidean CFT which is local, and hence does not agree with the usual
Euclidean CFT adjoint.  However, in \witc\ Witten introduced a
modified bulk inner product and corresponding adjoint.  We show that,
after a modification of the parity operation, Witten's bulk adjoint
induces precisely the standard non-local Euclidean CFT adjoint. We
further show that with the modified adjoint the $SL(2,C)$ generators
obey ${\cal L}_n^\dagger={\cal L}_{-n}$ (in a standard notation), as
opposed to the relation ${\cal L}_n^\dagger=\bar {\cal L}_{n}$ implied
by the naive adjoint.

As in the AdS case one expects that different coordinate systems in dS
are relevant for different physical situations.  In section 6 we
consider static coordinates for \ds, in which the metric is
\eqn\dscmt{{ds^2\over \ell^2}=-(1-r^2)dt^2 +
{dr^2 \over (1-r^2)}+r^2 d\varphi ^2,}
where $\ell$ is the de Sitter radius. 
These coordinates do not cover all of \ds\ with a single patch.
Nevertheless, they do cover the so-called southern diamond---the
region causally accessible to an observer at the `south pole' $r=0$.
Moreover, the symmetry generating time evolution of the southern
observer is manifest in static coordinates. Hence they appear
well-adapted to describing the physics accessible to a single
observer, as advocated in \klemm.
\im\ is at $r \to \infty$ and is conformal to a cylinder. 

In the $(t,r,\varphi)$ coordinates, the full \ds\ spacetime can be
covered with four patches separated by horizons.  We solve the scalar
wave equation in each patch and construct global solutions by matching
across the horizon. It is shown that the in vacuum on the cylinder and
the in vacuum on the sphere are equivalent.  A southern density matrix
is constructed from the Euclidean vacuum by tracing over modes which
are supported only in the northern causal diamond and are thereby
unobservable to the southern observer. This is explicitly shown to be
a thermal density matrix at temperature $T_{\rm dS}={1
\over 2
\pi\ell}$, with energy measured with respect to the static time
coordinate in \dscmt.  (This result is implicit in the original work
\gibbons.)

In section 7 we extend the static coordinate discussion to the
Kerr-\ds\ geometry which represents a pair of spinning point masses at
the north and south poles of \ds.  This has a Gibbons-Hawking
temperature $T_{GH}$ and angular potential $\Omega_{GH}$ which depend
on the mass and spin. It is shown that, after tracing over northern
modes, one obtains a thermal density matrix at precisely temperature
$T_{GH}$ and angular potential $\Omega_{GH}$.

According to the \ds /CFT$_2$ correspondence the quantum state on a
bulk spacelike slice ending on \im\ is dual to a CFT state on the
boundary of the spacelike slice at \im \dscft.  The dS-invariant bulk
vacuum should be dual to the $SL(2,C)$ invariant CFT vacuum.  
For pure de~Sitter space, we therefore expect to see a Casimir energy
$-c/12$, where $c={3\ell \over 2G}$ is the central charge of the CFT
computed in \dscft.  We find a two-parameter agreement with this
expectation by computing the Brown-York boundary stress tensor in
Kerr-\ds. This generalizes results of \klemm.

Finally, in section 8 we turn to the issue of de Sitter entropy.  In
the case of BTZ black holes in AdS$_3$, the entropy formula can be
microscopically derived, including the numerical coefficient, from the
properties of the asymptotic symmetry group together with the
assumption that the system is described by a consistent, unitary
quantum theory of gravity \asbh.  String theory seems necessary in
order to produce an actual example of such a theory, but the general
arguments follow from the stated assumptions independently of the
stringy examples.  Therefore it is natural to hope that a similar
discusion is possible for \ds.  We report here some partial results
but not a complete solution of the problem.  Related discussions appear in
\refs{\jmas,\banados,\park\bdm\ulf\cai\halyo{--}\hms}.
 
The main observation is that if we simply assume Cardy's formula for
the density of states, then a CFT with $c={3\ell \over 2G}$ at
temperature $T_{GH}$ and angular potential $\Omega_{GH}$ has a
microscopic entropy precisely equal to one quarter the area of the
Kerr-\ds\ horizon.  The two-parameter fit is striking but at present
should be regarded as highly suggestive numerology rather than a
derivation.  For one thing, the dual CFT is unlikely to be unitary
\dscft, and so there is no reason for Cardy's formula to
apply. Secondly, it is not clear how a mixed thermal state arises in
the dual CFT. The natural CFT state associated to \im\ is the $SL(2,C)$
invariant vacuum, in agreement with the pure nature of the global bulk
de Sitter vacuum. A mixed density matrix arises in the bulk only after
tracing over the unobservable northern modes. However, tracing over
northern modes is a bulk concept.  We have not succeeded in finding a
natural boundary interpretation of this operation.

We believe this raises a sharp and important 
question whose answer may lie within the present framework 
and in particular may not require 
a stringy construction of de Sitter.  What is the meaning, in terms of the 
dual boundary CFT, of tracing out  degrees of freedom which are 
inaccessible to a single observer? 

Two appendices detail useful properties of hypergeometric functions and 
de Sitter Green functions.  
For the rest of the paper we will set $\ell=1$ unless otherwise stated.

\newsec{Green Functions}
The two point Wightman function of a
free massive scalar can be used to characterize the various de Sitter
invariant vacua. In this section we describe these Green functions
and their properties. Previous studies of scalar field theory in 
de Sitter space, largely concentrating on the
four-dimensional case, can be found in 
\refs{\mottola,\allen,\cht\tagirov\figari\rumpf\abbott\najmi
\ford\allenfol\burges\figari{--}\birdav}.

\subsec{The Euclidean Vacuum and Wightman Function } In this
subsection we review the standard Euclidean vacuum and its
associated Wightman function.

$d$-dimensional de Sitter space (dS$_d$) is described by the
hyperboloid in $d+1$-dimensional Minkowski space
\eqn\hpr{P(X,X)=1,}
where
\eqn\nme{P(X,X^\prime)=\eta_{ab}X^aX^{\prime b},~~~~a,b=0,...,d.}
We will use lower case $x$ to denote a $d$-dimensional coordinate on
dS$_d$ and upper case $X$ to denote the corresponding
$d+1$-dimensional coordinate in the embedding space.  The function
$P(x,x')$ is greater than one for timelike separations, equal to one
for lightlike separations, and less than one for spacelike
separations.  In fact, $P(x,x')= \cos \theta $, where $\theta $ is the
geodesic distance between $x$ and $x'$ for spatial separations, or $i$
times the geodesic proper time difference for timelike separations.

A vacuum state $|\Omega \rangle$ for
a free massive scalar in de Sitter space with the mode expansion
\eqn\mex{\phi(x) = \sum_n \left[a_n \phi_n(x) + a_n^\dagger
\phi_n^*(x)\right]} can be defined by the conditions \eqn\ffl{
a_n|\Omega\rangle=0,} where $a_n$ and $a_n^\dagger$ as usual obey
\eqn\tyy{ [a_n,a^{\dagger}_m]=\delta_{nm}. } The modes $\phi_n(x)$
satisfy the de Sitter space wave equation \eqn\dddg{ (\nabla^2
-m^2) \phi_n=0,} and are normalized with respect to the invariant
Klein-Gordon inner product 
\eqn\rfv{ (\phi_n,\phi_m)=-i \int_\Sigma
d\Sigma^\mu\ \left( \phi_n \overleftrightarrow{\partial_\mu}
\phi_m^*\right)=\delta_{nm}.} 
The integral is taken over a complete spacelike slice $\Sigma$ in
dS$_d$ with induced metric $h_{ij}$, and $d\Sigma^\mu = d^dx \sqrt{h}
\, n^\mu$, where $n^\mu$ is the future directed unit normal vector.
The norm \rfv\ is independent of the choice of this
slice. $|\Omega\rangle$ depends on the choice of modes appearing in
\mex.

The Wightman function, defined by
\eqn\gcda{G_{\Omega}(x,x')=
\langle\Omega|\phi(x)\phi(x')|\Omega\rangle=\sum
_n\phi_n(x)\phi^*_n(x'),} characterizes the vacuum state
$|\Omega\rangle$.  There is a unique state, the ``Euclidean vacuum''
$|E\rangle$, whose Wightman function is obtained by analytic
continuation from the Euclidean sphere.  This state is invariant under
the full de Sitter group.  In $d$ spacetime dimensions the Wightman
function in the state $|E\rangle$ is
\eqn\gcd{\eqalign{G_{E}(x,x')&=\langle E|\phi(x)\phi(x')|E\rangle
= c_{m,d}F(h_+,h_-;{d\over 2}; {1+P(x,x') \over 2}),\cr
h_\pm &\equiv
{d-1 \over 2}\pm i \mu   \cr
\mu &\equiv \sqrt{m^2-\left({d-1\over2}\right)^2} \cr
 c_{m,d} &\equiv  
{\Gamma(h_+) \Gamma(h_-)\over(4 \pi)^{d/2} \Gamma({d\over 2})}.}}
$G_{E}$ is real in the spacelike region $P<1$ and singular on the 
light cone $P=1$. The $i\epsilon$
prescription near the singularity is
\eqn\singf{G_{E}(x,x') \sim \bigl( (t-t'-i\epsilon)^2- |\vec
x-\vec x'|^2)\bigr)^{1-{d \over 2}}.} Note that this prescription
cannot be written in terms of the invariant quantity $P$ alone,
which is time-reversal invariant. $G_E$ obeys
\eqn\ges{(\nabla^2-m^2)G_E(x,x')=0.}

In addition to the Wightman function, the Feynman propagator
\eqn\gfyn{G_F(x,x')=\Theta(t-t')G(x,x')+\Theta(t'-t)G(x',x)} and
commutator \eqn\gcm{G_C(x,x')=G(x,x')-G(x',x)} are also of
interest. With the normalization \gcd\ $G_F$ obeys
\eqn\gesf{(\nabla^2-m^2)G_F(x,x')={-i \over
\sqrt{-g}}\delta^d(x,x').}

\subsec{The MA Transform}

In this subsection we describe the MA (Mottola-Allen) transform
\refs{\mottola, \allen}, which relates the various de Sitter invariant
vacua and Wightman functions to one another.

Let $\phi^E_n(x)$ denote the positive frequency modes associated to
the Euclidean vacuum. Explicit expressions for $\phi^E_n$ will be
given later (sections 3.3 and 6.5), but we don't need them now. Let
$x_A$ denote the antipodal point to $x$ on the de Sitter hyperboloid
(i.e., $X_A=-X$). Then, as will be seen below, 
the Euclidean modes can be chosen to obey
\eqn\pfm{\phi^E_n(x_A)=\phi^{E*}_n(x).} Now consider a new set of
modes related by the MA transform\eqn\pfmz{\tilde \phi_n\equiv N_
\alpha (\phi^E_n +e^\alpha \phi^{E*}_n),~~~~N_\alpha \equiv{1
\over  \sqrt{1-e^{\alpha+\alpha^*}}}.}   
where $\a$ can be any complex number with ${\rm Re}\, \a < 0$.
The modes \pfmz\ can be used to define new operators $\tilde a_n$ and
$\tilde a_n^\dagger$ via a decomposition of the form \mex .  These are
related to the Euclidean operators $a^E_n$ and $a^E_n{}^\dagger$ by
\eqn\anew{
\tilde a_n = N_\a (a^E_n -e^{\a^*} a^E_n{}^\dagger )
.}
This may be rewritten as 
\eqn\arela{\tilde a_n = \cU \, a^E_n \, \cU^\dagger,}
where
\eqn\uis{\cU = {\rm exp} \left\{
\sum_n c \, (a^E_n{}^\dagger)^2 - \bar{c} \, (a^E_n{})^2\right\},~~~~~	
c(\alpha) = {1\over 4}\left(\ln\tanh {-{\rm Re}\, \a\over 2} \right) 
e^{-i\, {\rm Im}\, \a}.}
The vacuum state
\eqn\ais{\av=\cU \, \ev}
is annihilated by the $\tilde a_n$.  The operator $\cU$ is unitary, so
\ais\ is properly normalized.  In the quantum optics literature, $\av$
is known as a squeezed state.  Equation \ais\ may be formally
rewritten as
\eqn\aiss{\av = C\, 
{\rm exp} \left(\half e^{\a^*}(a^E_n{}^\dagger)^2\right)|E\rangle, } 
where $C$ is a constant.  Although this expression is not normalizable
(so $C$ is technically zero), it is often more convenient than \ais.

The Wightman function in the state $\av$ is
\eqn\gds{G_{\alpha }(x,x')=\sum _n\tilde \phi_n(x)\tilde
\phi^*_n(x').} Using \pfm\ and \pfmz\ this can be rewritten as a sum
over Euclidean modes, \eqn\gdca{\eqalign{G_{\alpha
}(x,x')=N_\alpha^2\sum_n\bigl[& \phi^E_n(x)\phi^{E*}_n(x')
+e^{\alpha+\alpha^*} \phi^E_n(x')\phi^{E*}_n(x) \cr
&~~+e^{\alpha^*}
\phi^E_n(x) \phi^{E*}_n(x'_A)+e^{\alpha}
\phi^E_n(x_A)\phi^{E*}_n(x')\bigr],}} and then evaluated as
\eqn\gdc{G_{\alpha }(x,x')=N_\alpha^2\bigl[
G_{E}(x,x')+e^{\alpha+\alpha^*} G_{E}(x',x)
+e^{\alpha^*}  G_{E}(x,x'_A)+e^{\alpha}G_{E}(x_A,x') \bigr].}
Hence it is easy to obtain the $|\alpha\rangle$ Wightman function
from the Euclidean one. Since these Wightman functions depend only on
the $SO(d,1)$ invariant quantity $P$ (away from the singularities)
this construction demonstrates the invariance of the
$|\alpha\rangle$ vacua under the connected part of the
de Sitter group.
Note however that if $\alpha $ is not real the
collection of modes \pfmz\ is not mapped into itself by $CPT$. Therefore
the $|\alpha\rangle$ vacua are $CPT$ invariant only for real $\alpha $.

Of course, since the commutator of two fields is a c-number, the 
commutator function $G_C$ must be the same in all vacua.
It is easy to check that the commutator constructed from the two point 
function \gdc\ has this property.

The Wightman function \gdc\ has several peculiarities.  Firstly,
there are antipodal singularities at $x'=x_A$. However such
antipodal points are separated by a horizon so this singularity is
not observable.  Secondly, the singularity at coincident points
has a negative frequency component coming from the second term in
\gdc\ (although the commutator is unaffected). This means that
for $e^\alpha \neq 0$ the vacuum state does not approach the usual
Minkowskian one even at distances much shorter than the de Sitter
radius. This ``unphysical'' behavior
was to be expected since the MA transform \pfmz\ involves
arbitrarily high-frequency modes. Despite these peculiarities we will
see that these vacua play an interesting role in the dS/CFT
correspondence.

\subsec{Analytic Continuation from AdS}

An alternate way to get a dS Green function is by double analytic
continuation from AdS via the hyperbolic
plane.\foot{See \refs{\hulla,\vijay,\mcn} for discussions.}  In fact, we shall
argue that this yields a Green function which differs from any of those
discussed in the previous subsection and therefore, as far as we know,
is not physically realizable as the Wightman function in any vacuum state.
Hence the dS/CFT correspondence is not in any precise sense that we know 
of the analytic
continuation of the AdS/CFT correspondence, and care must be taken in
extrapolating from the latter to the former.

AdS$_d$ has a unique $SO(d-1,2)$ invariant vacuum whose scalar Green
functions can be obtained as a sum over normalizable eigenmodes. The
wave equation allows two possible falloffs (fast and slow) at
infinity, but only the fast falloff appears in the Green function.
Double analytic continuation from AdS to dS will therefore yield a dS
Green function with only one of the two possible falloff rates (which
become complex conjugates for large enough $m$). This cannot be the
Euclidean dS Green function, as the latter involves both falloffs.
There is a vacuum $|\alpha \rangle$ whose Green function has the
required falloff\foot{It turns out to correspond to the in vacuum
discussed below.}. However from
\gdc\ we see that the Green function
for every state except $|E\rangle$ has a coincident point singularity
with a coefficient larger than that of $|E\rangle$ and
containing two terms with opposite-signed $i\epsilon$ prescriptions.
However double analytic continuation from AdS will yield a  coincident
point
singularity with a canonical coefficient and a single $i\epsilon$
prescription. Hence it yields a Green function which is not realized
as $\langle\alpha |\phi(x)\phi(x')|\alpha \rangle$ for any $\alpha $.

\subsec{Particle Detection}

In this subsection we discuss particle detection by a geodesic
observer in the $|\alpha \rangle$ vacua. We will find a thermal spectrum only
for the Euclidean vacuum.

Consider an Unruh detector moving along a timelike geodesic, which
couples to the field as
\eqn\unr{\int dt \, m(t) \, \phi(x(t))}
where $m(t)$ is an operator acting on the internal states of the detector
and the integral is over the proper time along the detector worldline.
Without loss of generality we may take the detector to be sitting
on the south pole.  Let's assume that the detector has a spectrum of
states $|E_i\rangle$ with energies $E_i$,
and define the matrix element $m_{ij}=\langle E_i|m(0)|E_j\rangle$.
In the vacuum state $\av$ the
transition rate between the states $|E_i\rangle$
and $|E_j\rangle$ may be evaluated in perturbation theory
(see, e.g. the review \lectures)
\eqn\prate{
\dot{P}_\a(E_i\to E_j) =
|m_{ij}|^2 \int_{-\infty}^\infty dt\, e^{-i\Delta E t}\,
G_{\alpha}\left(x(t),x(0)\right)
}
where $\Delta E = E_j-E_i$.

First, let us study particle production in the Euclidean vacuum.
For two timelike separated points $x$ and $x'$ we have
$P(x,x')=\cosh t$ and $P(x_A,x')= - \cosh t$, where $t$ is
the proper time between $x$ and $x'$.  We take $t$ to be positive
(negative) if $x$ is in the future (past) light cone of $x'$.
As a function of $t$,  the appropriate $i\ep$ prescription
for the Wightman function is
\eqn\dcx{G_{E}(x,x')
=G_E(t-i\epsilon)
}
indicating that for positive (negative) $t$ we
should go under (over) the branch cut from $P=1$ to $P=\infty$ in
\gcd. As a function in the complex $t$ plane $G_E$ obeys
\eqn\ghj{G_E(t)=G_E(-t-2\pi i).} \eqn\gxhj{G^*_E(t)=G_E({\bar
t}-2\pi i).}  To evaluate $G_E(x',x)$ we must take $t\to-t$
\eqn\zdcx{G_{E}(x',x)=G_E(-t-i\epsilon)=G_E(t+i\epsilon-2\pi i
).}
Similarly, we may evaluate
\eqn\zdx{G_{E}(x,x'_A)=G_{E}(x_A,x')=G_E(t-i\pi).}
The points $x$ and $x'_A$ are spacelike separated, so it is
not necessary to insert an $i\epsilon$.

Let us consider the example of $d=3$.  
As a function of $t$, the Green function \gcd\ has singularities 
at $t=n\pi i$ for all $n\ne -1$.   This may be seen from
the alternate form of the Green function presented in Appendix A.
Thus in the evaluating \prate\ we may deform the contour of integration
in the complex $t$ plane
\eqn\dft{\eqalign{
\int_{-\infty}^\infty dt e^{-i\Delta E t}G_E(t-
i\epsilon) &=e^{-\pi \Delta E }\int_{-\infty}^\infty dt
e^{-i\Delta E t}G_E(t- i\pi)\cr &=e^{-2\pi \Delta E
}\int_{-\infty}^\infty dt e^{-i\Delta E t}G_E(t- 2\pi i
+i\epsilon) .}}
The $e^{-\epsilon E}$ terms have been dropped.
Using \ghj\ and the second line of \dft\ we find that
the detector response rate \prate\ obeys
\eqn\jljl{
{\dot{P}_E(E_i\to E_j) \over\dot{P}_E(E_j\to E_i)} =
e^{-2\pi\Delta E}
}
in the Euclidean vacuum.
This is the condition of detailed balance for a thermal system
at the de Sitter temperature
\eqn\temp{T_{dS} = {1\over2\pi}.}

For a general vacuum state $\av$ we may use the
identities \dft\ to relate the integrals of all four terms in
\gdc.
We find
\eqn\gdcc{\int_{-\infty}^\infty dt e^{-i\Delta E t}
G_{\alpha}(t-i\ep)=N_\a^2|1+e^{\a+\pi \Delta E}|^2
\int_{-\infty}^\infty dt e^{-i\Delta E t}
G_{E}(t- i\epsilon )  .}
So the ratio \jljl\ becomes\foot{This expression was obtained for the 
case of a scalar with conformal mass
in \burges.}
\eqn\kfkf{
{\dot{P}_\a(E_i\to E_j) \over
\dot{P}_\a(E_j\to E_i)} =
e^{-2\pi\Delta E}\left|{1+e^{\a+\pi\Delta E}\over1+e^{\a-\pi\Delta E}}\right|^2
.}
We conclude that the detector response is not thermal.
In general the detector will not equilibrate.
Even though the ratio \kfkf\ is non-zero, we will see in the next section 
that there are vacua for which, in a certain sense, there is no 
particle creation. 

\newsec{The Sphere}

In this section we study scalar field theory on $dS_{d}$
in global coordinates $(\t,\Omega)$.
The metric is
\eqn\dsmet{{ds^2  }=-d\t^2 + \cosh^2\t\, d\Omega_{d-1}^2,}
where $d\Omega_{d-1}^2$ is the usual metric on $S^{d-1}$,
parameterized by the coordinates $\Omega$.  A important feature of
these coordinates is that they cover all of $dS_{d}$ and hence are
suited to a global description of the quantum state.

\subsec{Solutions of the Wave Equation}

In this subsection we find solutions to the massive wave equation
\eqn\kg{(\nabla^2 - m^2 )\phi=0.}
This differential equation is separable, with solutions
\eqn\sep{\phi = y_L(\t) Y_{Lj}(\Omega).}
The $Y_{Lj}$ are spherical harmonics on $S^{d-1}$ obeying
\eqn\sph{\nabla^2_{S^{d-1}} Y_{Lj}=-L(L+d-2) Y_{Lj}.}
Here $L$ is a non-negative integer and $j$ is a collective index 
$(j_1,\dots,j_{d-2})$.
We will use a non-standard choice of $Y_{Lj}$'s, with
\eqn\ylmnorm{
Y_{Lj} (\Omega_A) = Y^*_{Lj} (\Omega)
= (-)^L Y_{Lj}(\Omega).}
Here $\Omega_A$ denotes the point on $S^{d-1}$ antipodal to
$\Omega$.  In terms of the usual spherical harmonics $S_{Lj}$,
\eqn\ydef{
Y_{Lj} = 
\sqrt{i\over2} \, S_{Lj} + (-)^L \sqrt{-i\over2} \, S^*_{Lj}.
}
The functions $Y_{Lj}$ are orthonormal,
\eqn\krutz{\int d\Omega\, Y_{Lj}(\Omega) Y^*_{L'j'}(\Omega) =
\delta_{LL'} \delta_{jj'},}
and complete,
\eqn\krotz{\sum_{Lj} Y_{Lj}(\Omega) Y^*_{Lj}(\Omega') =
\delta^{d-1}(\Omega,\Omega').}

We then have
\eqn\yeq{\ddot y_L + (d-1) \tanh \t \dot y_L +
\left[ m^2  + {L(L+d-2)\over \cosh^2\t}\right] y_L = 0.}
In terms of the coordinate $\sigma=-e^{2\tau}$ this becomes
\eqn\yyeq{\sigma(1-\sigma)y_L'' + \left[ (1-{d-1\over2}) -
(1+{d-1\over2})\sigma\right] y_L' + \left[
{m^2  \over 4}{1-\sigma\over \sigma} - {L(L+d-2)\over1-\sigma}\right] y_L = 0.}
Let us make the substitution 
\eqn\skit{y_L^{\ri}=\cosh^L\t\, e^{(L+{d-1\over2}-i\mu)\t} x.}  
With
\eqn\jk{\mu=\sqrt{ m^2-{(d-1)^2\over4} },}
equation \yyeq\ becomes a hypergeometric equation for $x$,
\eqn\xeq{\sigma(1-\sigma) x'' + \left[c-(1+a+b)\sigma\right] x'- abx=0,}
with coefficients
\eqn\coff{a=L+{d-1\over2},~~~b=L+{d-1\over2}-i\mu,~~~c=1-i\mu.}

Let us consider the case of real positive $\mu$, i.e., $2m>(d-1)$.
We find that
\eqn\yp{{y^{\ri}_L}= {2^{L+d/2-1}\over\sqrt{\mu}}
\cosh^L\t e^{(L+{d-1\over2}- i\mu)\t}
F(L+{d-1\over2},L+{d-1\over2}- i\mu;1- i\mu;-e^{2\t})}
and its complex conjugate are two linearly independent solutions.
The normalization is fixed by demanding that these modes are
orthonormal with respect to the inner product \rfv, which is easily
evaluated on $\cal{I}^-$.

\subsec{In and Out Vacua}

We now use the solutions \yp\ to construct in (out) vacua with no
incoming (outgoing) particles, and find the Bogolyubov transformation
relating them.  Note that \yeq\ is invariant under time reversal.
Hence we obtain another pair of linearly independent solutions by
defining
\eqn\yre{
{y^{\ro}_L}(\t)={y^{\ri}_L}^*(-\t).}
Explicitly,
\eqn\yf{{y^{\ro}_L}= {2^{L+d/2-1}\over\sqrt{\mu}}
\cosh^L\t e^{(-L-{d-1\over2}- i\mu)\t}
F(L+{d-1\over2},L+{d-1\over2}+ i\mu;1+ i\mu;-e^{-2\t}).}

At the past boundary ($\t \to - \infty$) we find that $F\to 1$ and hence
\eqn\ypb{ {y^{\ri}_L} \to {2^{d/2-1}\over\sqrt\mu} 
e^{({d-1\over2} - i\mu)\t}}
while at the future boundary ($\t \to \infty$)
\eqn\yfb{ {y^{\ro}_L} \to  {2^{d/2-1}\over\sqrt\mu}
e^{-({d-1\over2} + i\mu)\t}.}  
Thus we see that the modes
\eqn\gmod{\eqalign{
\phi^{\ri }_{Lj} (x) &= {y^{\ri}_L} (\t) Y_{Lj} (\Omega) \cr
\phi^{\ro }_{Lj} (x) &= {y^{\ro}_L} (\t) Y_{Lj} (\Omega) .
}} 
are positive frequency modes with respect to the global time $\t$ near
the asymptotic past and future boundaries, respectively.  They
represent incoming and outgoing particle states.  They define two
vacua, $|\ri \rangle $ and $|\ro \rangle$, which are annihilated by
the lowering operators associated to $\phi^{\ri }$ and $\phi^{\ro }$,
respectively.  Physically, $|\ri \rangle $ is the state with no
incoming particles on
\im\ and $|\ro \rangle$ is the state with no outgoing particles on
\ip.

The Bogolyubov coefficients relating the two sets of modes can be found
by using the hypergeometric transformation equations (summarized in
Appendix B) and \ylmnorm.  One finds
%
\eqn\bdef{
\phi^{\ri }_{Lj} = A e^{-2i\theta_L} \phi^{\ro }_{Lj} 
+i B {\phi^{\ro }_{Lj}}^*.
}
where
\eqn\abis{
A=\left\{\eqalign{&1,~~~~~~~~~~~~{\rm d \ odd} \cr
           &\coth\pi\mu,~~~{\rm d \ even} }\right.
\ \ , \ \ \ \ \
B=\left\{\eqalign{&0,~~~~~~~~~~~~~~~~~~~~~{\rm d \ odd} \cr
           &(-)^{d\over2}\csch\pi\mu,~~~{\rm d \ even} }\right.
\ \ ;}
we have isolated the phase
\eqn\thetais{
e^{-2i\theta_L} =
(-)^{L-{d-1\over2}} {\G(-i\mu) \G(L+{d-1\over2}+i\mu)\over\G(i\mu)
\G(L+{d-1\over2}-i\mu)}
}
for later convenience.  The coefficients obey $|A|^2 - |B|^2 = 1$ as
required for properly normalized modes.

Note that $B$, the coefficient mixing positive and negative frequency
modes, vanishes in odd dimensions.  This implies that the two sets of
modes define the same vacuum:
\eqn\rtu{ |\ri \rangle = |\ro \rangle~~~~~{\rm in  \ odd \ dimensions}.}
Hence, there is no particle production.  If no particles are coming in
from \im, no particles will go out on \ip.\foot{Note however that according
to \kfkf\ an Unruh detector still observes particles.} 
This is in contrast to the
even-dimensional case for which there is always some particle
production.

From \ypb\ it follows that $\phi^{\ri }_{Lj} \sim e^{h_- \t}$ near
\im.  In the language of \dscft, this implies the modes $\phi^{\ri}$
are dual to operators of weight $h_+$ on the boundary.  Likewise,
$\phi^{\ri}{}^*$ are dual to operators of weight $h_-$.  The de Sitter
transformations act on the boundary theory as global conformal
transformations, which do not mix operators of different weight.  We
conclude that $\phi^{\ri}$ and $\phi^{\ri}{}^*$ do not mix under the
de Sitter group, so the states $\iv$ and $\ov$ are de Sitter
invariant.

It is convenient to define the rescaled global modes
\eqn\reg{\eqalign{
\tilde{\phi}^{\ri }_{Lj} (x) &=
e^{i\theta_L} {y^{\ri}_L} (\t) Y_{Lj}(\Omega) \cr
\tilde{\phi}^{\ro }_{Lj} (x) &=
e^{- i\theta_L} {y^{\ro}_L} (\t) Y_{Lj}(\Omega).
}}
This is a trivial phase shift, so
$|\ri \rangle$ and $|\ro \rangle$
are the states annihilated by the lowering operators associated to
$\tilde{\phi}^{\ri }$ and $\tilde{\phi}^{\ro }$, respectively.
In this basis the Bogolyubov transformation
\eqn\bog{
\tilde{\phi}^{\ri }_{Lj} (x) =
 A \tilde{\phi}^{\ro }_{Lj} (x) +i B {\tilde{\phi}^{\ro }_{Lj}}{}^* (x)
}
has the form of an MA transform, and so can be used to define additional
de Sitter invariant vacua.
The modes \reg\ have the useful property that for any point $x$
\eqn\rega{
\tilde{\phi}^{\ri }_{Lj} (x_A) = \tilde{\phi}^{\ro}_{Lj}{}^* (x)
}
where $x_A \sim (-\tau,\Omega_A)$ is the point antipodal to $x$.
In odd dimensions this becomes
\eqn\ant{
\tilde{\phi}^{\ri }_{Lj} (x_A) = {\tilde{\phi}^{\ri }_{Lj}}{}^* (x)
.}
This implies that in odd dimensions the $\ri$ vacuum is CPT invariant, 
whereas in even dimensions CPT interchanges $\ri$ and $\ro$.

\subsec{The Euclidean Vacuum}
In this subsection we construct the Euclidean vacuum $|E\rangle$ in the basis
of spherical modes.

The Lorentzian de~Sitter geometry \dsmet\ can be continued to
Euclidean signature by taking $\tau$ to run along the imaginary $\tau$
axis, from $\tau=-{i\pi\over 2}$ to $\tau={i\pi\over 2}$. The
resulting geometry is a round $d$-sphere.  We define the upper (lower)
Euclidean hemisphere as the portion of this path that lies in the
upper (lower) complex $\tau$ plane.  In particular, the upper (lower)
Euclidean pole lies at $\tau = {i\pi\over 2}$ ($\tau = -{i\pi\over
2}$).

We define positive frequency Euclidean modes to be those that are
regular when analytically continued to the lower Euclidean
hemisphere.  In this subsection we find these modes in global
coordinates.  The Euclidean vacuum $|E\rangle$ is the state that is
annihilated by the positive frequency Euclidean modes.

We may rewrite \yyeq\ in terms of the variable $z=1-\sigma=1+e^{2
\tau}$, which is well suited to analyzing the behavior of global modes
on the Euclidean geometry.  Upon substituting
\eqn\sket{y_L^E=\cosh^L\t\, e^{(L+{d-1\over2}+i\mu)\t} x,}
we obtain the hypergeometric equation  
\eqn\zeq{z(1-z) {d^2x\over dz^2}
 + \left[\hat{c}-(1+a+b^*)\sigma\right] {dx\over dz}- ab^* x=0,}
with positive integer coefficient
\eqn\cofz{\hat{c} = 2L+d-1.}
We find the general solution
\eqn\xxis{
x=C U_1 +D U_2}
where 
\eqn\punk{U_1= F(L+{d-1\over2},L+{d-1\over2}+i\mu;2L+d-1;z).}
The second solution is given by
\eqn\funk{U_2 = z^{2-2L-d}
F(1-L-{d-1\over2},1+i\mu-L-{d-1\over2};3-2L-d;z)}
if $d$ is odd, and by
\eqn\dunk{U_2 = U_1 \ln z + \sum_{k=2-2L-d}^\infty Q_k z^k}
if $d$ is even; the coefficients $Q_k$ are found, e.g., in \gradsh.

The Lorentzian geometry lies on the path from $z=1$ (${\cal I}^-$)
along the real $z$ axis to $z=+\infty$ (${\cal I}^+$).  On the throat,
at $z=2$, it intersects with the Euclidean geometry, which lies on a
unit circle centered at $z=1$.  The lower (upper) hemisphere
corresponds to the lower (upper) half-circle.  The Euclidean poles are
at $z=0$.  The functions \xxis\ have a branch cut from $z=1$ to
$z=+\infty$.  Hence, they are not analytic on the whole Euclidean
sphere.  By choosing the Lorentzian path to run just below the real
axis ($z \rightarrow z-i\epsilon$), we obtain solutions that are
analytic on the lower hemisphere and the entire Lorentzian
geometry.

The first solution \punk\ is regular in these regions, whereas the
second solution, \funk\ or \dunk, becomes singular at the lower
Euclidean pole, at $z=0-i\epsilon$.  Hence we discard the second set
of modes and keep the first.  The modes can be analytically continued
through the branch cut to the upper hemisphere, where they are not
expected to be regular.

The normalized Euclidean modes are
\eqn\neu{
\phi^E_{Lj} (x) =
{1\over f_L \sqrt{e^{2\pi \mu}-1}} y^E_L(\t) Y_{Lj} (\Omega)
}
where
\eqn\ye{\eqalign{
y^E_L &= {2^{L+d/2-1}i^{-L+{d-1\over2}}\over\sqrt{\mu}}
\cosh^L\t e^{(L+{d-1\over2}+ i\mu)\t} \cr
&~~~~~~~~~~~~~~~F(L+{d-1\over2},L+{d-1\over2}
+ i\mu;2L+d-1;1+e^{2\t}) \cr
f_L &= {\G (2L+d-1) \over \G(L+{d-1\over2})}
\left|{\G(i\mu)\over\G(L+{d-1\over2}-i\mu)}\right|
}}

The Euclidean Green function \gcd\ is then given by the mode sum
\eqn\ged{G_E(x,x')=\sum_{L,j}\phi^E_{Lj}(x)\phi^{E*}_{Lj}(x').}
This expression was given in the four-dimensional case in \mottola.

\subsec{The $|E\rangle \to |\ri \rangle$ Transformation}
In this subsection we show that the Euclidean and in vacua are MA
transforms of each other.

Let us again specialize to the case of $2m>(d-1)$.
The $y^E$ are then related to the $y^\ri$ by
\eqn\rel{
y^E_{L} = f_L \left( (-)^{L+{d-1\over2}} e^{-i\theta_L} {y^{\ri}_L}^*
+ e^{\pi \mu + i\theta_L} {y^{\ri}_L} \right).
}
So the Euclidean modes are related to the global modes by
\eqn\ebog{
\phi^E_{Lj} = {1\over\sqrt{1-e^{-2\pi\mu}}}  \left( \tilde{\phi}^{\ri }_{Lj}
    + (-)^{d-1\over2} e^{-\pi\mu}
    {\tilde{\phi}^{\ri }_{Lj}}{}^* \right).
}
from which it follows, along with \ant\ and \bog, that
\eqn\yant{
\phi^E_{Lj}(x_A)=\phi^E_{Lj}{}^*(x)
}
in any dimension.  This implies the Euclidean vacuum is CPT invariant.

Now, \ebog\ may be inverted to give
\eqn\all{
\tilde{\phi}^{\ri }_{Lj}
 = {1\over\sqrt{1-e^{-2\pi\mu}}} \left({\phi}^{E}_{Lj}
      + (-)^{{d+1\over2}} e^{-\pi\mu}{{\phi}^{E}_{Lj}}{}^* \right)
}
which is an MA transformation with
\eqn\ab{
\a = -\pi \mu +i\left({d+1\over2}\right)\pi .
} 
We have thus identified the MA transformation relating the $\iv$
vacuum and the Euclidean vacuum $\ev$.

\newsec{CFT Interpretation }

In this section we interpret the CPT invariant (real $\alpha $) family
of bulk de Sitter invariant vacua as a line of marginal deformations
of the boundary CFT. A similar interpretation may extend to the the
case of general complex $\alpha $ but we do not pursue it here.  In
this and later sections we restrict to the case $d=3$.

\subsec{ ${\cal I}^\pm$ Correlators}
In this subsection we evaluate the various Green functions appearing
on the right hand side of \gdc\ for $x$ and $x'$ on ${\cal I}^\pm$,
and then put the results together to see how the boundary values of
the correlators depend on $\alpha$.  We use global coordinates
$(\t,\s)$, $\s=(w,\bw)$, where $w=\tan {\theta\over 2} e^{i\varphi}$
is the complex coordinate on the 2-sphere, so that
\eqn\dsmnt{ds^2=-d\t^2+4\cosh^2\t{dwd \bw \over(1+w\bw)^2}.}

The behavior of the correlators at ${\cal I}^\pm$ follows from
the asymptotic form of the hypergeometric functions.
As $|z| \to \infty$ one has (see Appendix B)
\eqn\tzsf{\eqalign{
F(h_+,h_-;{3 \over 2};z) &\to c_+(-z)^{-h_+}+c_-(-z)^{-h_-},\cr
c_\pm&={\Gamma({3 \over 2}) \Gamma
(h_\mp-h_\pm) \over  \Gamma(h_\mp) \Gamma({3 \over 2}-h_\pm)}.
}}
This expression is not in general real
(unless $z$ is real and negative) because the $h_\pm =1\pm i\mu$ are not real.
In spherical coordinates one finds near \im\
\eqn\pxcf{\lim_{\tau,\tau'\to-\infty}P(\tau ,\s;\tau ', \s')=-
{e^{-\tau-\tau'}|w-w'|^2 \over 2(1+w\bar w)(1+ w' \bar w')}.}
For $x=(\t,\s)$ and $x'=(\t',\s')$ both on \im\
\eqn\rtv{\lim_{\tau ,\tau' \to -\infty}G_E(x, x')=
  e^{h_+(\tau+\tau')}\Delta_{+}(\s;\s')
 +e^{h_-(\tau+\tau')}\Delta_{-}(\s;\s').}
$\Delta_{\pm}$ here is proportional to the two point function for a conformal
field of dimension $h_\pm$ on the sphere:
\eqn\wwd{\Delta_{\pm}(\s;\s')=
4^{h_\pm}c_{m,d}c_\pm
\bigr[{ (1+w \bar w)(1+w' \bar w') \over  |w-w'|^2} \bigr]^{h_\pm}.}
We note that $G_E(x,x')=G_E(x',x)$ on \im\ as the points are spacelike
separated. We have assumed here, and in the following expressions
(unless explicitly stated) that $x$ and $x'$ are not coincident so that 
contact terms can be ignored. 

Let us now consider the case where $x$ is on \im\ and $x'$ is on \ip.
Since the antipodal point to $x'$, namely
$x'_A = (-\tau', \s'_A) = (-\tau', -{1 \over \bar w'},-{1 \over w'})$,
is on \im\
we may use \rtv\ and the formula \eqn\zpcf{P(x,x')=-P(x,x'_A).}
In continuing \rtv\ to positive $P$ we must take care to
go above the branch cut, in accord with the
$i\epsilon $ presription for the Wightman function with $\tau'>\tau$.
We find
\eqn\zztav{\lim_{\tau\to-\infty\atop\t'\to\infty}G_E(x,x')=-
 e^{h_+(\tau-\tau')}e^{-\pi \mu}\Delta_{+}(\s;\s'_A)
- e^{h_-(\tau-\tau')}e^{\pi \mu}\Delta_{-}(\s;\s'_A)
.}
To evaluate $G_E(x',x)$ we must go under the branch cut, yielding
\eqn\tazvd{\lim_{\tau\to-\infty\atop\t'\to\infty}G_E(x',x)=-
 e^{h_+(\tau-\tau')}e^{\pi \mu}\Delta_{+}(\s;\s'_A)
- e^{h_-(\tau-\tau')}e^{-\pi \mu}\Delta_{-}(\s;\s'_A).}

Now we insert these results into formula \gdc\ for the Wightman function
in the general vacuum state $|\alpha \rangle$. For both
points on \im\ one finds
\eqn\gab{\eqalign{
 \lim_{\tau,\tau' \to - \infty}
G_{\alpha }(x,x')
&= N_\alpha^2
(1-e^{\a+\pi\mu})(1-e^{\a^*-\pi\mu})
e^{h_+(\tau+\tau')}\Delta_{+}(\s;\s')\cr
& + N_\alpha^2
(1-e^{\a-\pi\mu})(1-e^{\a^*+\pi\mu})
e^{h_-(\tau+\tau')}\Delta_{-}(\s;\s').}}
On the other hand for $x$ on \im\ and $x'$ on \ip\ we get
\eqn\gzab{\eqalign{\lim_{\tau\to-\infty\atop\t'\to\infty}
G_{\alpha }(x,x')
=& -N_\alpha^2
e^{-\pi\mu}|1-e^{\a+\pi\mu}|^2
e^{h_+(\tau-\tau')}\Delta_{+}(\s;\s'_A)\cr
& -N_\alpha^2
e^{\pi\mu}|1-e^{\a-\pi\mu}|^2
e^{h_-(\tau-\tau')}\Delta_{-}(\s;\s'_A).}}
We see that the boundary correlators depend nontrivially on the
choice of vacuum. Since we have taken $|P| \to \infty$, these formulae are
valid only for non-coincident points on ${\cal I}^\pm$ and omit possible
contact terms.

Let us now turn to the interesting special case of the $\ri$ vacuum,
which has $\alpha=-\pi \mu$.
For both points on \im\  it follows from \gab\
that the correlators vanish! 
On the other hand, for $x$ on \im\ and
$x'$ on \ip\ we get
\eqn\flu{\eqalign{\lim_{\tau\to-\infty\atop\t'\to\infty}
G_\ri (x,x')
&=-2\sinh \pi\mu\, e^{h_-(\tau-\tau')}\Delta_{-}(\s;\s'_A)
,\cr
\lim_{\tau\to-\infty\atop\t'\to\infty}G_\ri (x',x)
&=-2\sinh \pi\mu\, e^{h_+(\tau-\tau')}\Delta_{+}(\s;\s'_A).}}

When the points on \im\ coincide there is a
contact term which can be easily computed by noting that
the Wightman function on \im\ reduces to a mode sum
over spherical harmonics. This gives
\eqn\dll{\lim_{\tau,\tau' \to -\infty}G_{\rm in}(x,x')=
{2\over\mu} 
e^{h_-\t+h_+\t'}\delta^2(\s,\s').}

The situation can be described as follows. As \im\ is approached, the
spatial part of the scalar kinetic terms are exponentially supressed
relative to the rest of the action. Neighboring points decouple and
the theory becomes ultralocal.  It reduces to a harmonic oscillator at
each point; hence the vanishing of $G_\ri$. 
However, the map defined by propagation from \im\ to \ip\ is
not ultralocal on the sphere.  It introduces nontrivial correlators
when one point is on \im\ and the other is on \ip.

Of course, in other vacua---such as the Euclidean vacuum---there are
nontrivial \im\ correlators.  As will be seen in the next subsection,
the wave functions for these vacua differ from the in vacuum
wavefunction by terms which are nonlocal on \im. These terms are
directly responsible for the nontrivial \im\ correlators.

\subsec{dS Vacua as Marginal CFT Deformations}

Now we argue that the dual interpretation of the one-parameter family of
\ds\ vacua is a one-parameter family of marginal deformations of the
CFT.  It is convenient to define operators on \im\ and \ip\ by
\eqn\rfvv{\eqalign{
\lim_{\t \to-\infty}\phi(\tau,\s)&=\phi^\ri_+(\s)e^{h_+ \tau}+\phi^{\ri
}_-(\s) e^{h_-\t},\cr\lim_{\t \to
\infty}\phi(\tau,\s_A)&=\phi^\ro_+(\s)e^{-h_+ \tau}+\phi^{\ro
}_-(\s) e^{-h_-\t}.\cr}}
$\phi^\ro_\pm$ has been defined with an antipodal inversion relative 
to $\phi^\ri_\pm$ so that they transform the same way under conformal
transformations \dscft.
These are position space versions of the
creation operators associated to the spherical
modes $\phi^{\ri}$ and $\phi^{\ro}$, 
\eqn\modrel{\eqalign{
\phi^\ri_+(\s) = (\phi^\ri_-(\s))^\dagger
&= \sqrt{2\over\mu} \sum_{Lj}a^\ri_{Lj}{}^\dagger \, Y_{Lj}^*(\s) \cr
\phi^\ro_+(\s) = (\phi^\ro_-(\s))^\dagger
&= \sqrt{2\over\mu} \sum_{Lj}a^\ro_{Lj}{} \, Y_{Lj}(\s_A)
}}
From the asymptotic Green functions \dll\ and \flu\ we find that the only
non-zero commutators are
\eqn\pll{\eqalign{[\phi^\ri_-(\s),\phi^{\ri}_+(\s')]
&=[\phi^\ro_+(\s),\phi^{\ro}_-(\s')]=
{2\over\mu}\delta^2(\s,\s'),\cr
[\phi^\ri_\pm(\s),\phi^{\ro}_\pm(\s')]&= 
\pm 2 \sinh \pi\mu\,\Delta_\pm(\s,\s').}}

The in and out operators are related by a Bogolyubov
transformation and hence are not independent. In this subsection
we take $\phi^\ri_\pm$ to be the fundamental operators.
At a general point in the bulk $\phi$ is determined from
its value on \im\ via
\eqn\lftr{\phi(x)=i\int_{{\cal I}^-}d^2x'\sqrt{g}\, 
G_C(x,x')\overleftrightarrow \p_{\t'} \phi(x').}
In particular, taking $x$ to be on \ip\ and using the limiting
expression for $G_C$ (which does not depend on $\alpha $) we find
\eqn\pot{\phi^\ro_\pm(\s) = 
-\mu \sinh \pi \mu \int d^2 \s' \Delta_\pm(\s,\s') \,
\phi^\ri_\mp(\s').}
This is a position-space version of the Bogolyubov transformation
\bdef.\foot{In fact, expression \pot\ is singular for $\s=\s'$ and so
is really defined by \bdef.}  We see that the absence of mixing
between $\phi^\ri$ and $\phi^{\ro *}$, which seemed so surprising in
section 3.2, follows directly from the asymptotic behavior of the
Green functions. We note parenthetically that this implies the
identity (verified in \idet)
\eqn\rfgl{ (\mu \sinh \pi \mu )^2
\int d^2 \s'' \Delta_-(\s,\s'')\Delta_+(\s'',\s')=\delta^2(\s,\s').}

The $\iv$ vacuum obeys
\eqn\trin{\phi^\ri_-(\s)\iv=0.}
The general $\av$ vacuum state discussed in section 2.2
can be constructed in terms of the in vacuum as
\eqn\tll{\av=\exp\left\{
c(\gamma) {\mu\over 2}\int d^2\s\, \phi^\ri_+\phi^\ro_-
-c(\bar{\gamma}) {\mu\over 2} \int d^2\s\, \phi^\ri_-\phi^\ro_+
\right\}\iv, }
where 
\eqn\gma{
e^\gamma = {\sinh {\pi\mu+\alpha\over 2}
\over \sinh {\pi\mu-\alpha\over2}},}
and the function $c$ is given by \uis.  This equation may be formally
rewritten as
\eqn\tlla{\av = C \exp \left(e^{\gamma^*} 
{\mu\over 4}\int d^2\s\, \phi^\ri_+\phi^\ro_- \right) \iv.}
These vacua obey the manifestly $SL(2,C)$ invariant condition
\eqn\trzin{ 
\phi^\ri_-(\s)\av = -e^{\gamma^*}  \phi^\ro_-(\s)\av   .}
This is most easily seen by applying the representation $\phi^\ri_- =
- {2\over\mu} {\delta\over\delta\phi^\ri_+}$ of \pll\ to \tlla.
In particular, the Euclidean vacuum has
$\alpha =-\infty$ and therefore obeys
\eqn\trin{ \phi^\ri_-(\s)\ev = e^{-\pi \mu} \phi^\ro_-(\s)\ev    .}

Now we consider the boundary field theory. Consider the two
operators $\opm$  dual to 
$\phi^\ri_\mp$ 	
with conformal
weights $h_\pm$. According to the dS/CFT correspondence
\dscft, the dual $\opm$ correlators are determined from the
$\phi^\ri_\pm$ correlator \gab\ as
\eqn\ddl{ \langle\alpha |\opm(\s) \opm(\s') | \alpha\rangle=
-{\mu^2\over 4} N_\alpha^2(1-e^{\a\pm\pi\mu})(1-e^{\a^*\mp\pi\mu})
\Delta_\pm(\s,\s').}
The commutators \pll\ also imply the contact terms
\eqn\cnt{\eqalign{
\langle \alpha |\om(\s) \op(\s') | \alpha \rangle
&={1 \over 1-e^{\gamma+\gamma^*}}\,{\mu\over 2}\delta^2(\s,\s'), \cr
\langle \alpha |\op(\s) \om(\s') | \alpha \rangle
&={e^{\gamma+\gamma^*}
\over 1-e^{\gamma+\gamma^*}}\,{\mu\over 2}\delta^2(\s,\s').}}
From the CFT point of view this is an unusual contact term prescription in
that it depends on the operator ordering. 

What is the CFT origin of the parameter $\alpha$? Usually a
one-parameter family of correlators corresponds to a line of marginal
deformations generated by a dimension $(1,1)$ operator. Indeed, $\op
\om$ is a dimension $(1,1)$ operator. Let us consider adding this
operator to the two-dimensional CFT action with real coefficient
$\lambda$.  At linear order this perturbs the correlators according to
the formula
\eqn\dlp{\eqalign{\delta_{\lambda}
\langle\alpha |\op(\s) \op(\s') | \alpha\rangle
&= -\langle\alpha | {\lambda\over 2} \int d\s'' \left\{ \op(\s'') \om(\s''),
\op(\s) \op(\s') \right\} |\alpha\rangle \cr
&= -  4\mu \lambda \coth {\gamma} \,
 \langle\alpha |\op(\s) \op(\s') | \alpha\rangle.}}
%
Let's take $\a$ to be real, so that $\langle\alpha |\op\op|
\alpha\rangle$ is a monotonically increasing function of $\alpha$.
Then the variation of the two point function as $\a\to\a+\ep$ is
proportional to the deformation \dlp, which may be integrated to
determine $\lambda$ as a function of $\a$.  

This strongly suggests that the
family of $CPT$ and $SO(d,1)$ invariant vacuum states are marginal
deformations of the boundary CFT generated by the $(1,1)$ operator
$\op \om$. The two point functions of these CFTs can all be made
equivalent by rescaling operators, except for the special case $\alpha
=-\pi \mu$. So in principle from this analysis alone the CFTs with
$\alpha\ne\-\pi\mu$ might all be equivalent. In order to complete 
the argument one should check that the three point function is not
invariant under such rescalings. This has been shown 
in \nm.\foot{We thank Greg Moore for
discussions on this point.} 

\newsec{$\cc \cp \ct$ and the Inner Product}

In this section we discuss various choices of norm for the Hilbert
space of a real scalar field on \ds, or equivalently the definition of
the adjoint.  The first naive choice one might make is
\eqn\rtd{\phi^\dagger(x)=\phi(x).}
However Witten \witc\ argues that this choice may not be well-defined
for full quantum gravity oustide of perturbation theory. An alternate
norm is proposed \witc\ which involves path integral evolution form
\im\ to \ip\ together with $\cc \cp \ct$ conjugation. In this section
we will explicitly compute this norm for a free scalar and find, after
a slight modification involving the form of $\cp$, that it has a very
natural boundary interpretation: it yields the Zamolodchikov metric
for the boundary CFT.

Before delving into details it is instructive to recall an isomorphic
discussion of norms which arises in the standard treatment of
Euclidean CFT.  Consider the mode expansion for a free boson on the
Lorentzian cylinder (ignoring zero modes)
\eqn\fwz{X(\sigma^+,\sigma^-)=i\sum_{n}\bigl( 
{\alpha_{n} \over n}e^{-2\pi in\sigma^+}  +{\bar
\alpha_{n} \over n}e^{2\pi in\sigma^-}  \bigr ).}
Using $\alpha_{-n}^\dagger=\alpha_{n}$ one finds 
\eqn\fwbz{X^\dagger(\sigma^+,\sigma^-)=X(\sigma^+,\sigma^-).}
On the other hand, the standard mode expansion on the complex Euclidean
plane is
\eqn\febz{X(z, \bar z)=i\sum_{n}{1\over n}\bigl( {\alpha_{n} \over z^n}
+{\bar\alpha_{n} \over\bar z^n}\bigr )}
Using $\alpha_{-n}^\dagger=\alpha_{n}$ one now finds 
\eqn\fcz{\eqalign{X^\dagger(z, \bar z)&=-i\sum_{n}{1\over n}\bigl( 
{\alpha_{-n} \over \bar z^n}+{\bar
\alpha_{-n} \over z^n}\bigr)\cr 
& =X({1 \over \bar z}, {1 \over z}).}}  
In this case the adjoint relates $X$ at points in the Euclidean plane
reflected across the unit circle. In particular the norm of the state
created by $X(z, \bar z)$ or any other operator is just the two point
function, and hence is the Zamolodchikov norm.

Returning now to \ds, the naive adjoint rule \rtd\ induces an adjoint
in the Euclidean boundary CFT of the form $X^\dagger(z, \bar z)= X (z,
\bar z) $. On the other hand we will show that the modified Witten
adjoint gives precisely \fcz.  We further consider the \ds\ $SL(2,C)$
isometry generators ${\cal L}_n,~ \bar {\cal L}_n$, for $n=0,\pm 1$.
It is shown that ${\cal L}_n^\dagger= \bar {\cal L}_n$ for the naive
adjoint, but ${\cal L}_n^\dagger= {\cal L}_{-n}$ for the modified
adjoint.

Although we take $d=3$, much of the following discussion carries over
simply to higher dimensions.

\subsec{Continuous and Discrete Symmetries of de Sitter Space}
 
\ds\ can be represented by the hyberboloid
\eqn\dsh{X^+X^- + z \bar z=\ell^2}
in flat Minkowski space.   The isometries of \ds\ are then inherited from
the $SL(2,C)$ Lorentz isometries of Minkowski space.  The six generators can be
written as  combinations $\vec J+i\vec K$ of rotations and boosts
together with their complex conjugates.  We denote the associated Killing
vectors by $\zeta_n$ and
$\bar \zeta_n$ for $n=0,\pm 1$.
The past and future horizons of an observer worldline at $z=0$
are located on the hyperboloid at $X^+X^-=0$. We denote the Killing vectors
preserving this horizon as
\eqn\loo{\eqalign{\zeta_{0}+\bar \zeta_{0}&= X^+\p_+-X^-\p_-,\cr
\zeta_{0}-\bar \zeta_{0}&= \bar z \p_{\bar z}-z\p_z.}}
The four additional Killing vectors are
\eqn\loco{\eqalign{\zeta_{1} &= X^+\p_z- \bar z \p_-,\cr
 \zeta_{- 1} &= X^-\p_{\bar z}-  z \p_+ , \cr
\bar \zeta_{1} &= X^+\p_{ \bar z}-  z \p_- ,\cr
\bar \zeta_{- 1} &= X^-\p_{ z}- \bar z \p_+ . \cr
}}
They obey the Lie bracket relation
\eqn\fdt{[\zeta_m,\zeta_n]=(n-m)\zeta_{m+n}.}

In addition, we consider the two discrete symmetries parity and time
reversal
\eqn\tlg{\eqalign{PX^\pm&=X^\pm,~~~~~Pz=-{ z},\cr
TX^\pm&=X^\mp,~~~~~Tz=z.}}
In terms of the global coordinates $(\t,\Omega)$,
$P$ takes a point $\Omega=(\theta,\varphi)$ on the 2-sphere to the point
$P\Omega=(\theta,\pi+\varphi)$
and $T$ takes $\tau$ to $-\tau$.

Our choice of parity $P$ in \tlg\ reflects all the coordinates about
an observer at the south pole. An alternate choice is $Pz=\bar z$
which reflects only one coordinate. This is the choice employed in
\witc, motivated by the fact that the corresponding $\cc\cp\ct$
operation is known to be an exact field theory symmetry, after taking
a flat space limit of \ds. We shall indicate below how the results are
modified if this definition of $P$ is employed.

\subsec{$\cc\cp\ct$}
We now compute the action of the discrete symmetries
$\cc$, $\cp$ and $\ct$ on the field operators. 

We consider a real scalar field so that $\cc$ is trivial. 
We wish to find Hilbert space operators $\cp$ and $\ct$ that implement
\tlg\ on $\phi(x)$ as \eqn\ptr{\cp \phi(x)\cp=\phi(Px),~~~~~~~~\ct
\phi(x)\ct=\phi(Tx).}  As usual $\ct=UK$ is an antilinear operator
which combines a unitary operator $U$ with complex conjugation $K$
of functions. 

The mode expansions for $\phi$ in terms of the $\phi^\ri$ and $\phi^\ro$
modes are
\eqn\serp{\phi(\t,\s)
=\sum_{L,j}\bigl( a^\ri_{Lj}y_L^\ri(\t) Y_{Lj}(\Omega)+
b^\ri_{Lj}y_L^{\ri *}(\t) Y^*_{Lj}(\Omega)\bigr)}
\eqn\serq{\phi(\t,\s)
=\sum_{L,j}\bigl( a^\ro_{Lj}y_L^\ro(\t) Y_{Lj}(\Omega)+
b^\ro_{Lj}y_L^{\ro *}(\t) Y^*_{Lj}(\Omega)\bigr)}
We have written lowering and raising operators as $a$'s and
$b$'s, respectively, and are not assuming here that $a^\dagger=b$.

We define the action of $\cp$ by
\eqn\cpd{ \cp
a^\ri_{Lj}\cp = (-)^{j}a^\ri_{Lj},~~~~~~~~~~~~
  \cp  b^\ri_{Lj} \cp = (-)^{j} b^\ri_{Lj}}
and similarly for the out operators. 
Since $Y_{Lj}(P\Omega) = (-)^j Y_{Lj}(\Omega)$ 
this definition reproduces \ptr. 
We define the action of $\ct$ by
\eqn\cpzd{ \ct a^\ri_{Lj}\ct =(-)^L a^\ro_{Lj},~~~~~~
  \ct  b^\ri_{Lj} \ct =(-)^L b^\ro_{Lj}.}
At the same time it acts as complex conjugation on 
functions. The wave functions appearing in \serp\ 
transform as 
\eqn\frtl{\eqalign{Y^*_{Lj} (\Omega) &= (-)^L Y_{Lj}(\Omega),\cr
{y^{\ri}_L}^*(\t)&={y^{\ro}_L}(-\t).}}
Putting this together gives \eqn\srp{\eqalign{\ct \phi(\t,\s) \ct
&=\sum_{L,j}\bigl( a^\ro_{Lj}y_L^\ro(-\t) Y_{Lj}(\Omega)+
b^\ro_{Lj}y_L^{\ro *}(-\t) Y^*_{Lj}(\Omega)\bigr)\cr
&=\phi(-\t,\s) ,}}
as required. 

We wish to consider the action of $\cc\cp\ct$ on the in and out field 
operators $\phi_+^\ri$ and $\phi_+^\ro$ defined by \rfvv.
Using \cpd --\frtl\ these obey 
\eqn\dstx{\eqalign{\cp\ct \phi_+^\ri(\s)\cp\ct &=\phi_-^\ro(P\s_A)\cr
\cp\ct \phi_+^\ro(\s)\cp\ct &=\phi_-^\ri(P\s_A).}}

\subsec{The Witten Inner Product and Modifications}

Following Witten \witc, we now describe a modified inner product.
First we construct a bilinear pairing between states on \im and states
on \ip.  We will consider asymptotic states on \ipm\
\foot{ 
Of course, these states are linear in $b^\ri$ and $b^\ro$.   
The general asymptotic states will take a more complicated form.
}
\eqn\ios{|\Psi^\ri\rangle=\int \Psi^\ri(\s)\phi_+^\ri(\s) |\ri\rangle,
~~~~~~
|\Psi^\ro\rangle=\int \Psi^\ro(\s)\phi_-^\ro(\s) |\ri\rangle,} 
where $\Psi^\ri(\Omega)$ and $\Psi^\ro(\Omega)$ are functions on the
2-sphere.  Using \pot, the out state can be expressed as a linear
combination of in states
\eqn\ios{
|\Psi^\ro\rangle=-\mu\sinh\pi\mu \int\int
 \Psi^\ro(\s')\Delta_-(\s',\s)\phi_+^\ri(\s)|\ri\rangle .}  
This corresponds to evolving the state $|\Psi^\ro\rangle$ backwards
from \ip\ to \im, and defines the bilinear pairing
\eqn\dzz{(\Psi^\ro|\Psi^\ri)=-\mu\sinh\pi\mu
\int\int\Psi^\ro(\s')\Delta_-(\s',\s)\Psi^\ri(\s).} 
We now use the pairing to define an inner product on \im\ that 
is antilinear in the first argument.  Note that
applying $\cc\cp\ct$ to a state on \im\ gives us a state on \ip:
\eqn\gklre{
\cc\cp\ct|\Psi^\ri\rangle=
\int \Psi^{\ri}{}^*(P\s_A)\phi_-^\ro(\s) |\ri\rangle}
to which we may apply the pairing \dzz.  We find the inner product
between two states on \im
\eqn\wipp{\langle \Psi^\ri|\Upsilon^\ri\rangle =-\mu\sinh\pi\mu
\int\int\Psi^{\ri *}(\Omega) 
\Delta_-(P\Omega_A,\Omega')\Upsilon^\ri(\Omega').}
For free field theory the norm \wipp\ implies the 
adjoint relations\foot{It is intriguing that this adjoint relates
degrees of freedom separated by a horizon.}
\eqn\wadj{\eqalign{\phi^\dagger(x)&=\phi(PTx),\cr
\phi^{\ri\dagger}_\pm(\Omega)&=\phi^\ro_\pm (P\Omega_A)   .}}

This may look strange at first but is in fact precisely the usual norm
employed  for a Euclidean CFT. 
Note that $P$ coupled with the antipodal map is reflection about
the equator, so that 
\eqn\dcv{PA(z, \bar z)= ({1 \over \bar z}, {1 \over z}),}
as in \fcz. For states constructed by acting with operators on 
\im, it therefore follows that the norm is simply the two point function. 
Hence \wipp\ gives the Zamolodchikov metric on the boundary CFT.

Formula \wipp\ in fact remains valid for any choice of $P$. Using $Pz
=\bar z$ as in \witc, one finds instead of \dcv, $PA(z, \bar z)= (-{1
\over z}, -{1 \over \bar z})$. The adjoint then involves rotation by
$\pi$ about $z=\pm i $ rather than reflection across the unit disc.

\subsec{Adjoints of the $SL(2,C)$ Generators}

The quantum generators of the symmetries \loo\ and \loco\ are as
usual given by 
\eqn\lnm{\eqalign{{\cal L}_n&=\int_\Sigma d\Sigma^\mu
{\cal T}_{\mu\nu}\zeta_n^\nu,\cr 
\bar {\cal L}_n &=\int_\Sigma d\Sigma^\mu
{\cal T}_{\mu\nu}\bar \zeta_n^\nu ,}}
for any complete spacelike slice $\Sigma$.
We choose $\Sigma$ to be 
the throat $X^+=X^-$ because it is mapped to itself under
both $P$ and $T$.  For a massive scalar,
\eqn\hnm{{\cal T}_{\mu\nu}(x)=\p_\mu \phi(x)
 \p_\nu \phi(x) -\half g_{\mu\nu}\left[ (\nabla \phi(x))^2+
m^2\phi^2(x)\right].}
With the ordinary inner product, ${\cal T}_{\mu\nu}$
is hermitian, and one finds ${\cal L}_n^\dagger = \bar {\cal L}_n$.
With the modified inner product, one has
\eqn\hnm{{\cal L}_n^\dagger=
\int d\Sigma^\mu (x) {\cal T}_{\mu\nu}(PTx)\bar \zeta_n^\nu(x).}
We then consider a coordinate transformation $x'=PTx$. One finds
\eqn\hnm{{\cal L}_n^\dagger=
\int d\Sigma^\mu(PTx') 
{\cal T}_{\mu\nu}(x')\bar \zeta_n^\nu(PTx').}
Using the relations
\eqn\dsgc{\eqalign{ \zeta_n(PTx') &=  - \bar \zeta_{-n}(x'),\cr
d\Sigma^\mu(PTx') &= -d\Sigma^\mu(x'),}}
it follows that
\eqn\lkz{{\cal L}_n^\dagger = {\cal L}_{-n}.}

In \dscft\ the $SL(2,C)$ isometries of \ds\ were conjectured to extend to 
a full Virasoro symmetry of the full quantum gravity (not just a free
scalar). This naturally acted not on closed spacelike
slices but on asymptotically flat slices ending on $\cal I$. 
It would be interesting to compute the adjoints of these generators.

\newsec{The Cylinder}

In this section we study scalar field theory in static
coordinates. Again for simplicity we specialize to
\ds, although we expect the higher dimensional cases to be similar.
The metric is
\eqn\dscmet{{ds^2}=-(1-r^2)dt^2 +
{dr^2 \over (1-r^2)}+r^2 d\varphi ^2.}
This metric is singular at the horizons $r=1$, which divides dS$_3$
into four regions.  There are two regions with $0\leq r<1$
corresponding to the causal diamonds of observers at the north and
south poles. We shall refer to these as the northern and southern
diamonds. There are two more regions with $1<r < \infty$ containing
\ip\ and \im\ which we shall refer to as the future and past triangles.
On ${\cal I}^\pm$, where $r \to \infty$, the spatial metric approaches
$r^2(dt^2+d\varphi^2)$ and hence is conformal to the cylinder.

Unlike the global coordinates, static coordinates do not smoothly
cover all of dS$_d$. However, they are well-suited to describing the
physics associated to an observer who can access a single causal
diamond. The Killing vector ${\p \over \p t}$ is manifest in static
coordinates, but is future-directed only in the southern diamond; it
is past-directed in the northern diamond and space-like in the past
and future triangles.  In the following we solve the scalar wave
equation in the four regions.  Then we patch the solutions together to
get a global solution over all of \ds\ by matching at the horizons. We
further show explicitly that tracing the Euclidean vacuum over the
Hilbert space of the northern modes leads to a thermal density matrix
in the southern diamond.

\subsec{The Wave Equation}

The equation of motion for a scalar field of mass $m$ is
$(\nabla^2-m^2)\phi=0$.  In static coordinates, this
becomes
\eqn\steom{\left[-{1\over 1-r^2} \p_t^2
+ {1\over r}\p_r (1-r^2) r\p_r +
{1\over r^2} \p_\varphi^2 - m^2\right] \phi =0.}
The equation separates, so that a general solution can be expanded
\eqn\modes{\phi(t,r,\varphi) = \int_0^\infty d\omega
\sum_{j=-\infty}^\infty a_{\omega j} \phi_{\omega j}  +
b_{\omega j} \phi^{\omega j} + a^\dagger_{\omega j} \phi^*_{\omega j}
+ b^{\dagger}_{\omega j} \phi^{\omega j*},}
where
\eqn\hghh{\phi_{\omega j} = f_{\omega j}(r)
e^{-i\omega t+ij\varphi},~~~~
\phi^{\omega j} = f^{\omega j}(r) e^{-i\omega t+ij\varphi},}
and $f_{\omega j}(r)$, $f^{\omega j}(r)$ are two linearly independent
solutions of the radial equation
\eqn\radi{(1-r^2) {d^2f_{\omega j}\over dr^2}
 + \left({1\over r} - 3r \right)
{df_{\omega j}\over dr} + \left( {\omega^2\over 1-r^2} 
- {j^2\over r^2} - m^2
\right) f_{\omega j} =0.}

\subsec{The Northern and Southern Diamonds}

A solution smooth near $r=0$ is given by
\eqn\forig{\eqalign{\phi^{S}_{\omega  j} &= f_{\omega j}(r)
e^{-i\omega t+ij\varphi},\cr
f_{\omega j}(r)&\equiv
r^{|j|}
(1-r^2)^{i\omega\over 2} F(a,b;c;r^2), \cr
a &\equiv \half (|j|+i\omega+h_+ ), \cr
b&\equiv \half (|j|+i\omega+h_- ), \cr
c & \equiv 1+|j|.
}}
We have not normalized this solution, although the necessary factor
follows from computations below.  The superscript $S$ denotes that
this solution is in the southern diamond.
One can show from the transformation formulae for hypergeometric
functions (see Appendix B) that
\eqn\gji{f^*_{\omega j}=f_{-\omega j}=f_{\omega j}.}
Similarly we may define northern modes
\eqn\frigb{\phi^{N}_{\omega  j} = f_{ \omega j}(r)
e^{-i\omega t+ij\varphi}.}
It is convenient to use the time coordinate $t$ both in the northern
and in the southern diamond.  Although this coordinate system does not
uniquely label points on all of \ds, there will be no confusion since
we denote northern functions with a superscript $N$.  The coordinate
$t$ runs forward in the southern diamond and backward in the northern
diamond.  Hence for $\omega>0$ the modes \forig\ are positive
frequency and \frigb\ are negative frequency.

Near the horizon, for $r\rightarrow 1$, one can show (see Appendix B
for details) that \forig\ becomes:
\eqn\fhxddr{\eqalign{\phi^{S}_{\omega  j} &\to
e^{-i\omega t+ij\varphi}\Gamma(1+|j|)
\bigl[{\Gamma
(-i\omega) \over  \Gamma(\half(|j|-i\omega +h_+) )\Gamma(\half( |j|
-i\omega +h_-))
}(1-r^2)^{i\omega\over 2}\cr &~~+
{\Gamma
(i\omega) \over  \Gamma(\half(|j|+i\omega +h_+)) \Gamma(\half(|j|+
i\omega +h_-))  }
(1-r^2)^{-{i\omega\over 2}}\bigr]
.}}In order to analyze the flux across the horizons it is useful to
introduce
Kruskal coordinates
\eqn\ksdc{\eqalign{r&={1+UV \over 1-UV},\cr
t&=\half \ln(-{U \over V}),\cr}}
in which
\eqn\dscl{ds^2={1 \over( 1-UV)^2}\bigl(-4dUdV+(1+UV)^2 d \varphi^2 \bigr).}
$U>0$ and $V<0$ in the southern diamond. The future (past) horizon
is at  $V=0$ ($U=0$). In contrast to the static
coordinates, Kruskal coordinates are nonsingular at the horizon.

The modes \fhxddr\ become, for $r\to 1$ ($UV \to 0$):
\eqn\zsx{\phi^S_{\omega  j} \to
e^{ij\varphi}
\bigl[\alpha_{\omega j}
(-V)^{i\omega}+\alpha^{*}_{\omega j}
U^{-{i\omega }}\bigr]
,}
where we define the complex constants
\eqn\adef{\alpha_{\omega j}\equiv {\Gamma(1+|j|)\Gamma
(-i\omega)2^{i\omega}\over \Gamma(\half(|j|-i\omega +h_+)
)\Gamma(\half(|j| -i\omega +h_-)) }=\alpha^{*}_{-\omega j} .}
The first term in \zsx\ is incoming flux across the past horizon,
while the second is outgoing flux across the future horizon. A similar
analysis in the northern diamond with $U<0,~V>0$ gives for $r \to 1$:
\eqn\nxr{\phi^N_{\omega  j} \to e^{ij\varphi}\bigl[\alpha_{\omega j}
V^{i\omega}+\alpha^*_{\omega j} (-U)^{-i\omega }\bigr] .}
The northern and southern modes are simply related by
\eqn\xr{\phi^S_{\omega  j}(-U,-V) = \phi^{N}_{\omega  j}(U,V).}

The second family of  solutions is given by
\eqn\sorig{\phi^{\omega j} = \ln(r^2) \phi_{\omega j} +
e^{-i\omega t+ij\varphi} r^{|j|}(1-r^2)^{i\omega\over 2}
\sum_{n=-|j|}^{\infty} A_n r^{2n},}
where the coefficients $A_n$ are given in, e.g., equation 15.5.19 of
\abramowitz.  These modes are singular at r=0 for all $j$ and 
hence are excluded.

\subsec{The Past and Future Triangles}

Let us analyze the behavior of the modes in the past triangle (which
includes ${\cal I}^-$ but not ${\cal I}^+$) where $r^2>1$.  A complex
solution of \steom\ is
\eqn\fdrig{\eqalign{\phi^{\ri +}_{\omega  j} &= f^+_{\omega j}(r)
e^{-i\omega t+ij\varphi},\cr f^+_{\omega j}(r)&\equiv r^{-h_+} (1-{ 1
\over r^2})^{i\omega\over 2} F(a,1-a^*;h_+;{ 1 \over r^2} ).}}
%
Using properties of the hypergeometric functions
one finds that $f^+_{\omega j}$ is invariant under
$\omega \to -\omega$, but is not real. Therefore
the second solution of \steom\ is obtained by complex conjugation:
\eqn\ssol{ \phi^{\ri -}_{\omega  j} = (f^+_{\omega j}(r))^*
e^{-i\omega t+ij\varphi}.}
This is equivalent to replacing $h_+$ with $h_-$ in \fdrig.
Near \im\ we find 
\eqn\pbin{
\phi^{\ri \pm}_{\omega j} \sim r^{-h_\pm}.
}
In the past triangle the coordinate $r$ is timelike and past-directed,
so that the $\phi^{\ri -}$ are positive frequency for $m^2>1$. 

Near the horizon, for $r\rightarrow 1$, we find
\eqn\fhxr{\eqalign{\phi^{\ri +}_{\omega  j} &\to
e^{-i\omega t+ij\varphi}\Gamma(h_+)
\bigl[{\Gamma
(-i\omega) \over  \Gamma(\half(|j|-i\omega +h_+) )\Gamma(\half(-|j|
-i\omega +h_+))
}(r^2-1)^{i\omega\over 2}\cr &~~+
{\Gamma
(i\omega) \over  \Gamma(\half(-|j|+i\omega +h_+)) \Gamma(\half(|j|+
i\omega +h_+))  }
(r^2-1)^{-{i\omega\over 2}}\bigr]
.}}

The relation between static and
Kruskal coordinates in the past triangle is
\eqn\ksc{\eqalign{r&={1+UV \over 1-UV},\cr
t&=\half \ln({U \over V}).\cr}}
$U$ and $V$ are both negative in this region. The boundary
with the northern (southern) diamond is at
$V=0$ ($U=0$). The near horizon behavior \fhxr\ becomes
\eqn\fhxdr{\phi^{\ri +}_{\omega  j} \to
e^{ij\varphi}
\bigl[\beta_{\omega  j}
(-V)^{i\omega}+\beta_{-\omega  j}
(-U)^{-{i\omega }}\bigr]
,}
where
\eqn\dft{\beta_{\omega  j}\equiv {\Gamma(h_+)\Gamma
(-i\omega)2^{i\omega}\over  \Gamma(\half(|j|-i\omega +h_+) )\Gamma(\half(-|j|
-i\omega +h_+))}.}
Similarly one finds near $r=1$ that
\eqn\xdr{\phi^{\ri -}_{\omega  j} \to
e^{ij\varphi}
\bigl[\beta^*_{-\omega  j}
(-V)^{i\omega}+\beta^*_{\omega  j}
(-U)^{-{i\omega }}\bigr]
.}

One may also define modes in the future triangle by
\eqn\fdrie{\eqalign{\phi^{\ro +}_{\omega  j} &= f^+_{\omega j}(r)
e^{-i\omega t+ij\varphi},\cr
\phi^{\ro -}_{\omega  j} &= (f^+_{\omega j}(r))^*
e^{-i\omega t+ij\varphi}.}}
Near \ip\ we find 
\eqn\pbout{
\phi^{\ro \pm}_{\omega j} \sim r^{-h_\pm}.
}
In the future triangle the coordinate $r$ is future-directed,
so that the $\phi^{\ro +}$ are positive frequency. 

The relation between static and Kruskal coordinates in the future
triangle is again given by \ksc, which means that $t$ increases to the
south (north) in the future (past) triangle.  $U$ and $V$ are both
positive in this region. The boundary of the future triangle with the
northern (southern) diamond is at $U=0$ ($V=0$).  Near the horizons
($UV=0$) the $\phi^\ro$ modes obey
\eqn\ddr{\eqalign{\phi^{\ro +}_{\omega  j} &=
e^{ij\varphi}
\bigl[\beta_{\omega  j}
V^{i\omega}+\beta_{-\omega  j}
U^{-{i\omega }}\bigr],\cr
\phi^{\ro -}_{\omega  j}& =
e^{ij\varphi}
\bigl[\beta^*_{-\omega  j}
V^{i\omega}+\beta^*_{\omega  j}
U^{-{i\omega }}\bigr]
.}}
The past and future modes are simply related by
\eqn\czm{\phi^{\ro \pm}_{\omega  j}(U,V)=\phi^{\ri \pm}_{\omega  j}(-U,-V).}

\subsec{Matching Across the Horizon.}

In the previous two subsections we have described solutions in the
past and future triangles as well as the northern and southern
diamonds. By matching fluxes across the horizon, these may be extended to
global solutions over all of \ds.  For example
the $(-V)^{i\omega}$ ($(-U)^{-i\omega}$) terms in the past modes
\fhxdr\ and \xdr\
carry flux into the southern (northern) diamond.  The continuation of
\fhxdr\ and \xdr\ into these regions is obtained by matching to \zsx\
along $U=0$ and to \nxr\ along $V=0$. Matching across the horizon
again then yields the future mode.

Henceforth we shall use the symbol $\phi^{\ri\pm}$ to denote the
global solution so constructed. Similarly, $\phi^{\ro\pm}$ will denote
the global solution agreeing with \ddr\ in the future triangle.  We
may also construct global solutions $\phi^S$ ($\phi^N$) that agree
with the modes \forig\ (\frigb) in the southern (northern)
diamond---these solutions vanish in the northern (southern) diamond.

From the matching procedure outlined above we find
that these modes obey
\eqn\fty{
\left(\eqalign{&\phi^S_{\omega j} \cr &\phi^N_{\omega j}}\right)
= {\bf A}_{\omega j}
\left(\eqalign{&\phi^{\ri+}_{\omega j} \cr &\phi^{\ri-}_{\omega j}}\right)
= {\bf A}^*_{\omega j}
\left(\eqalign{&\phi^{\ro-}_{\omega j} \cr &\phi^{\ro+}_{\omega j}}\right),}
where
\eqn\esz{N_{\omega j} {\bf A}_{\omega j}=
\left(\eqalign{\alpha_{\omega j}\beta^*_{\omega j}~~~~~~~
            & -\alpha_{\omega j}\beta_{-\omega j} \cr
              -\alpha^*_{\omega j}\beta^*_{-\omega j}~~~~~
            &  ~~~~\alpha^*_{\omega j}\beta_{\omega j}}    \right)}
and 
\eqn\gkfa{
N_{\omega j} \equiv 
(\beta_{\omega j}\beta^*_{\omega j}-\beta_{-\omega j}\beta^*_{-\omega j})
=-{\mu\over\omega}.  
}
Reversing the signs of $U$
and $V$ and using $\sigma_x {\bf A}\sigma_x={\bf A}^*$, one finds that
the second equation in~\fty\ follows from the first.  The Bogolyubov
transformation from ${\cal I}^-$ to ${\cal I}^+$ then follows from
\fty\ as
\eqn\ftxy{
\left(\eqalign{&\phi^{\ro-}_{\omega j} \cr &\phi^{\ro+}_{\omega j}}\right)
= {\bf B}_{\omega j}
\left(\eqalign{&\phi^{\ri+}_{\omega j} \cr &\phi^{\ri-}_{\omega j}}\right),}
where 
\eqn\esz{{\bf B}_{\omega j}=
\sigma_x  {\bf A}_{\omega j}^{-1}\sigma_x {\bf A}_{\omega j}
=\left(\eqalign{ {\alpha_{\omega j} \beta^*_{\omega j}
\over \alpha^*_{\omega j} \beta_{\omega j}}
         & ~~~~~~0\cr
             0~~~~~~
   & {\alpha^*_{\omega j} \beta_{\omega j}
\over \alpha_{\omega j} \beta^*_{\omega j}} }\right).}
As with the spherical modes of section 3.2, 
the Bogolyubov transformation \ftxy\
is trivial.  The vacuum $\iv$ defined by the modes
$\phi^{\ri}$ is identical to the vacuum $\ov$ defined by the
$\phi^{\ro}$.


%

\subsec{Euclidean Modes on the Cylinder }

In this subsection, following \unruh\
we write the Euclidean modes as linear combinations
of northern and southern modes.

In Kruskal coordinates the southern modes \forig\ in the southern diamond
are of the form
\eqn\rkm{\phi^S_{\omega j}=f_{\omega  j}(UV) e^{ij\varphi}
(-{V \over U})^{i \omega \over 2}}
for $U>0, V<0$, and vanish for $U<0, V>0$.
The  northern modes in the northern diamond are of the same form
\eqn\rckm{\phi^N_{\omega j}=f_{\omega  j}(UV) e^{ij\varphi}
(-{V \over U})^{i \omega \over 2} ,}
but have support for $U<0, V>0$ instead of $U>0, V<0$.  We wish to
find a linear combination of \rkm\ and \rckm\ which is analytic in the
lower complex $U$ and $V$ planes.\foot{Euclidean modes were defined
earlier to be regular on the lower Euclidean hemisphere ($\tau^{\rm
Re} =0$, $-{\pi\over 2} \leq \tau^{\rm Im} \leq 0$).  Explicit
transformation of coordinates shows that ${\rm sgn}~U^{\rm Im} ={\rm
sgn}~V^{\rm Im} = {\rm sgn}~\tau^{\rm Im}$.  The lower pole,
$\tau=-i{\pi\over 2}$, maps to a single point, $U=V=-i$, independently
of $\theta$.  Smooth curves through this pole remain smooth in the $U$
and $V$ planes.  Thus, modes that are analytic and bounded in the
lower half $U$ and $V$ planes will be regular on the lower Euclidean
hemisphere.}  This can be accomplished by analytically continuing the
southern modes~\rkm\ to the northern diamond along the contour
\eqn\fbb{U \to e^{-i\gamma}U,~~~~V \to e^{i\gamma}V ,}
taking $\gamma $ from $0$ to $\pi$. Notice that the product
$UV$ is independent of $\gamma $, so that the continuation of
the southern mode \rkm\ is
\eqn\rzkm{e^{ -\pi \omega}f_{\omega  j}(UV) e^{ij\varphi}
(-{V \over U})^{i \omega \over 2}.}
Comparing with \rckm\ we see that the linear combination
\eqn\lcm{\phi^E_{\omega j}=\phi^S_{\omega j}+ e^{ -\pi \omega}
\phi^N_{\omega j}}
is analytic in the lower half of the complex $U$ and $V$ planes. 
Since $t$ runs backwards
in the northern diamond, this is a linear combination
of positive and negative frequency modes.  A second linear combination
\eqn\lcms{\phi^{E\prime}_{\omega j}=(\phi^{N}_{\omega j})^*
+ e^{ -\pi \omega}(\phi^{S}_{\omega j})^*}
is also analytic in the lower half plane.
Both $\phi^E$ and $\phi^{E\prime}$ are positive frequency for
$\omega>0$.  


\subsec{MA Transform to Euclidean Modes}

In this subsection we will show that the $\iv$ vacuum on the cylinder
is the same as the $\iv$ vacuum on the sphere by showing that it is an
MA transform of the Euclidean vacuum with $\alpha =-\pi\mu$. This
result is anticipated by the fact that the dual CFTs should be simply
related by the conformal transformation from the sphere to the
cylinder.  Nevertheless, it provides a useful check on our
constructions.

The first step is to redefine $\phi^{\ri\pm}$ in order to simplify the
expression for ${\bf A }$ in \fty. Let
\eqn\llj{\eqalign{
\tilde \phi^{\ri+}_{\omega j}&=i^{j}{\alpha_{\omega j}\beta^*_{\omega
j}\over N_{\omega j}} \phi^{\ri+}_{\omega j}, \cr
\tilde \phi^{\ri-}_{\omega j}&=(-i)^{j}{\alpha^*_{\omega j}\beta_{\omega
j}\over N_{\omega j}} \phi^{\ri-}_{\omega j}.
}}
%
%
Then \fty\ becomes
\eqn\fty{
\left(\eqalign{&\phi^S_{\omega j} \cr &\phi^N_{\omega j}}\right)
=(-i)^j\left(\eqalign{~1~~~ & ~~~~q~~~ \cr ~(-)^j q & ~~(-)^j~}\right)
\left(\eqalign{&\tilde\phi^{\ri+}_{\omega j} \cr
               &\tilde\phi^{\ri-}_{\omega j}}\right),}
%
with
\eqn\aesz{q\equiv (-)^{j+1}\,
{\alpha_{\omega j}\beta_{-\omega j}\over
\alpha^*_{\omega j}\beta_{\omega j}}=
-{(-)^j+ e^{\pi(\omega+\mu)} \over e^{\pi\omega}+(-)^j e^{\pi\mu}}}
%
It follows that the Euclidean modes obey
\eqn\efc{\eqalign{\phi^E_{\omega j}&=
\phi^S_{\omega j}+e^{-\pi \omega}  \phi^N_{\omega j}\cr
&=
(-i)^j
{e^{\pi \omega}- e^{-\pi\omega} \over
e^{\pi \omega} + (-)^j e^{\pi \mu} }
\bigl(\tilde \phi^{\ri+}_{\omega j}-e^{\pi \mu}\tilde \phi^{\ri-}_{\omega
j}\bigr).}}
Inverting this relation, one recovers $\alpha =-\pi\mu$.

\subsec{The Thermal State}

Let us summarize the southern and northern mode expansions:
\eqn\nsme{\eqalign{
\phi^S(t,r,\varphi) = &\int_0^\infty d\omega \sum_{j=-\infty}^{\infty}
a^S_{\omega j} \phi^S_{\omega j} +
(a_{\omega j}^S)^\dagger (\phi^S_{\omega j})^* \cr
\phi^N(t,r,\varphi) = &\int_0^\infty d\omega \sum_{j=-\infty}^{\infty}
a^N_{\omega j} \phi^N_{\omega j} +
(a_{\omega j}^N)^\dagger (\phi^N_{\omega j})^*.}}
Here we take the modes $\phi^S$ and $\phi^N$ to be normalized with 
respect to the Klein-Gordon inner product \rfv. 
The Fock space in the southern diamond is constructed with lowering
operators $a^S_{\omega j}$ and raising operators $(a_{\omega
j}^S)^\dagger$.  The Fock space in the northern diamond is constructed
with lowering operators $(a_{\omega j}^N)^\dagger$ and raising
operators $a^N_{\omega j}$.  

The modes \lcm\ and \lcms\ annihilate the Euclidean vacuum, $|E\rangle$.
This allows us to express $|E\rangle$ as a superposition of states in
the northern and southern Fock spaces~\fiopre:
\eqn\etons{\eqalign{
|E\rangle & = \prod_{\omega=0}^\infty \prod_{j=-\infty}^\infty
\left( 1-e^{-2\pi\omega}
\right)^{1\over 2} \exp \left[ e^{-\pi\omega}  (a_{\omega j}^S)^\dagger
a_{\omega j}^N \right] |S\rangle \otimes |N\rangle \cr
 &= \prod_{\omega, j} \left( 1-e^{-2\pi\omega}
\right)^{1\over 2} \sum_{n_{\omega j}=0}^\infty 
e^{-\pi\omega n_{\omega j}} 
|n_{\omega j}, S\rangle \otimes |n_{\omega j}, N\rangle
.}}
Here $|S\rangle$ and $|N\rangle$ are the southern and northern vacua, and
\eqn\manny{\eqalign{
|n_{\omega j}, S\rangle= & (n_{\omega j}!)^{-{1\over 2}}
[(a_{\omega j}^S)^\dagger]^{n_{\omega j}} |S\rangle,\cr
|n_{\omega j}, N\rangle= & (n_{\omega j}!)^{-{1\over 2}}
[a_{\omega j}^N]^{n_{\omega j}} |N\rangle.}}
Only the southern diamond is causally accessible to an observer at the
south pole.  The quantum state in this region is described by a
density matrix $\rho^S$, which is obtained from a global state by
tracing over the field modes in the northern diamond.  For the
Euclidean vacuum \etons\ we obtain
\eqn\rhos{\rho^S_E = \tr_N |E\rangle\langle E|
= \prod_{\omega, j} \left[ \left(
1-e^{-2\pi\omega} \right) \sum_{n_{\omega j}} e^{-2\pi\omega
n_{\omega j}} |n_{\omega j}, S\rangle \langle n_{\omega j}, S| \right].}

Recall that the Killing vector $\xi^\mu \p_\mu= \p_t$ is everywhere
time-like and future directed in the southern diamond.  
Neglecting gravitational
back-reaction of the field modes, this allows us to define a
Hamiltonian for the southern modes:
\eqn\ham{{\cal M} = 
\int_{\Sigma^S} d\Sigma^\mu
{\cal T}_{\mu\nu} \xi^\nu 
= \int_0^\infty d\omega \sum_{j=-\infty}^{\infty}
(a_{\omega j}^S)^\dagger a_{\omega j}^S~ \omega,}
where ${\cal T}$ is the stress tensor of the scalar field.  Here
$\Sigma^S$ is a $t={\rm constant}$ Cauchy surface in the southern diamond 
with normal vector is $n_\Sigma^\mu\p_\mu =
(1-r^2)^{-\half}\p_t$.  This definition of energy is natural for the
observer at the south pole.  For later use, we also define the angular
momentum ${\cal J}$ as the conserved charge associated with the
Killing vector $\upsilon^\mu\p_\mu = -\p_\varphi$:
\eqn\ang{{\cal J} = 
\int_{\Sigma^S} d\Sigma^\mu
{\cal T}_{\mu\nu} \upsilon^\nu
= \int_0^\infty d\omega \sum_{j=-\infty}^{\infty}
(a_{\omega j}^S)^\dagger a_{\omega j}^S~ j.}

With respect to the Hamiltonian ${\cal M}$, the southern state
\rhos\ becomes a thermal density
matrix
\eqn\rhoth{\rho^S_E = C \exp\left(-{{\cal M}\over T}\right)}
with temperature $T={1\over 2\pi}$;  $C=\prod( 1-e^{-2\pi\omega})$ is
a normalization factor.

\newsec{Kerr-de~Sitter}

In this section we generalize the discussion of the previous
sections to the three-dimensional Kerr-de~Sitter solution,
which represents a spinning point mass in \ds.

\subsec{Static Coordinates}

The Kerr-de~Sitter metric describes the gravitational field of a point
particle whose mass and spin are parametrized by $1-M$ and $J$:
\eqn\stk{{ds^2} = -N^2 d t^2 + N^{-2} d r^2 +
 r^2 \left(d \varphi + N^{ \varphi} d t \right)^2.}
%
The lapse and shift functions are
\eqn\lsh{N^2 = M - r^2 +
{16 G^2 J^2 \over r^2},~~~~~~ N^{ \varphi} = -{4GJ \over r^2
}.}
The lapse function vanishes for one positive value of $r$:
\eqn\rplus{ r_+ = {1\over 2}\left( \sqrt{\tau} + \sqrt{\bar\tau}
 \right).}
where
\eqn\thlb{\tau \equiv M+i(8GJ).}
%
%
This is the cosmological event horizon surrounding an observer at
$r=0$.  It has a Bekenstein-Hawking entropy \refs{\bek,\hawk} of
\eqn\sbh{S = {\pi r_+ \over 2G}
= {\pi \over 4G} \left( \sqrt{\tau} + \sqrt{\bar\tau} \right).}
%
\subsec{Kerr-\ds\ as a Quotient of \ds}

In 2+1 dimensions, there is no black hole horizon for Kerr-de~Sitter
because the ``black hole'' degenerates to a conical singularity at the
origin.  This is best seen by writing the metric as an identification
of de~Sitter \deja.  Let us define $\mu = r_+$ and $\alpha = 4GJ/r_+$,
so that
\eqn\mjma{M = \mu^2 - \alpha^2,~~~J= {\mu\alpha \over 4G}.}
%
The coordinate transformation
\eqn\ctr{\eqalign{%
    \tilde t    &= \mu t    + \alpha \varphi, \cr
    \tilde \varphi &= \mu \varphi - \alpha {t}, \cr
    \tilde r    &= \sqrt{{r^2+ \alpha^2 \over \mu^2+\alpha^2}} \cr}}
changes the Kerr-de~Sitter metric to the vacuum form
%
\eqn\stds{{ds^2} = - \left(1 - {\tilde r^2} \right) d\tilde
t^2 + {d\tilde r^2
\over 1 - {\tilde r ^2}} + \tilde r ^2 d\tilde \varphi^2,}
but with a non-standard coordinate identification.  In empty de~Sitter
space, $(\tilde t,~ \tilde r ,~ \tilde \varphi+ 2\pi n)$ labels the same
point for all integer $n$.  In the presence of a particle, the points
\eqn\idbulk{(\tilde t,~ \tilde r,~ \tilde \varphi)
+ 2\pi n (\alpha,~ 0,~ \mu)}
are identified instead.

\subsec{Kerr-\ds\ Temperature and Angular Potential}

In this subsection we consider a scalar field in Kerr-\ds.  The
cylinder mode solutions found for de~Sitter in Section 6 are also
solutions in Kerr-de~Sitter, after the substitutions $t\rightarrow
\tilde t$, $r\rightarrow \tilde r$ and $\varphi \rightarrow \tilde
\varphi$ are performed.  For the modes to remain single-valued, the
angular momentum $j$ must be non-integer:
\eqn\jni{j = {n+\omega\alpha\over\mu},~~~~n~{\rm integer}.}
The mode analysis carries over trivially.  In particular, the
Euclidean modes \lcm\ and \lcms\ take the same form in Kerr-de~Sitter.

Analogues of \ham\ and \ang\ define conserved charges associated with
the Killing vectors $\tilde\xi^\mu \p_\mu= \p_{\tilde t}$ and
$\tilde\upsilon^\mu \p_\mu= -\p_{\tilde\varphi}$:
\eqn\haman{\eqalign{
\tilde {\cal M} = 
\int_{\Sigma^S} d\Sigma^\mu{\cal T}_{\mu\nu} \tilde\xi^\nu
& = \int_0^\infty d\omega \sum_{j=-\infty}^{\infty}
(a_{\omega j}^S)^\dagger a_{\omega j}^S~ \omega, \cr
\tilde {\cal J} = 
\int_{\Sigma^S} d\Sigma^\mu
{\cal T}_{\mu\nu} \tilde\upsilon^\nu 
& = \int_0^\infty d\omega \sum_{j=-\infty}^{\infty}
(a_{\omega j}^S)^\dagger a_{\omega j}^S~ j,}}
where ${\cal T}_{\mu\nu}$ is the matter stress tensor.
Here the hypersurface $\Sigma^S$ is defined, for example, by the
normal vector
\eqn\nyrp{n^\mu_{\Sigma^S} \p_\mu = 
{\tilde{r}\over\sqrt{1-\tilde{r}^2}} {\mu\over r} \p_{\tilde t}
+ {\sqrt{1-\tilde{r}^2}\over\tilde{r}} {\alpha\over
r}\p_{\tilde\phi}.}
(For $\alpha>0$, $\Sigma^S$ is not a space-like surface near the
origin; this does not affect the definition of conserved quantities.)
The expressions for $\tilde {\cal M}$ and $\tilde {\cal J}$ nevertheless
take the same form as ${\cal M}$ and ${\cal J}$ in de~Sitter space.
The Euclidean state, restricted to the southern diamond ($\tilde
r<1$), is a density matrix
\eqn\rhkds{\rho_E^S = C\exp\left(-{2\pi \tilde {\cal M}}\right).}

In the $(t,r,\varphi)$ coordinates, the asymptotic metric of Kerr-de
Sitter space takes a standard form near ${\cal I}$ (detailed in
section 7.4 below).  In order to compare conserved quantities of
different space-times, we must use the Killing vectors $\p_t$ and
$\p_\varphi$ to measure energy and angular momentum.\foot{We are
choosing the normalization of the time-like Killing vector to be fixed
at ${\cal I}$, as is appropriate for a CFT description.  By
normalizing at $\tilde{r}=0$ instead, one would obtain the apparent
temperature seen by a local observer
\bouhaw.}  The corresponding conserved charges are related to $\tilde
{\cal M}$ and $\tilde {\cal J}$ by a linear transformation.  Using
\ctr\ one finds
\eqn\thh{\tilde {\cal M} = {\mu\over\mu^2+\alpha^2} {\cal M} +
{\alpha\over\mu^2+\alpha^2} {\cal J}.}
Thus we obtain a density matrix
\eqn\rhkds{\rho^S_E = C \exp\left(-{{\cal M}+\Omega
{\cal J}\over T}\right),}
at temperature and angular potential
\eqn\tap{T = {\mu^2+\alpha^2\over 2\pi\mu},~~~~~
\Omega={\alpha\over\mu}.}

For later convenience it is useful to rewrite the
the density matrix \rhkds\ in terms of
the complex inverse temperature
\eqn\ctpp{\beta \equiv {1+i\Omega\over T} = 
{2\pi\over\sqrt{\bar{\tau}}},}
and the complex charges
\eqn\ccl{{\cal L}_0=\half({\cal M}- i{\cal J}),~~~~~\bar
{\cal L}_0=\half({\cal M}+ i{\cal J}).}
These charges are constructed from the 
complex Killing vector fields
\eqn\qzck{\zeta_0 = \half (\p_t + i \p_\varphi),~~~~~
\bar \zeta_0 = \half (\p_t - i \p_\varphi).}
Then the density matrix of the scalar field in the southern diamond
takes the form
\eqn\rtxn{\rho^S_E=C \exp\left(-\beta {\cal L}_0 -\bar \beta \bar
{\cal L}_0 \right).}

\subsec{The Boundary Stress Tensor and Virasoro Charges}

In this subsection we define, compute and interpret the Brown-York
boundary
stress tensor in static coordinates, following~\klemm.

In the static coordinates ${\cal I}^\pm$ is at $r\to\infty$.  The metric takes the asymptotic form
\eqn\rfzl{ ds^2=-{dr^2 \over r^2}+({r^2}
-{M \over 2})dwd\bar w
+{\tau \over 4}dw^2 +{\bar \tau \over 4}d\bar w^2+{\cal O}({1 \over
r^4}),}
with
\eqn\bbh{w\equiv \varphi +i {t}.}
Since $w \sim w+2\pi $, the boundary is a cylinder with conformal
metric
\eqn\dddflc{ds_{\rm conf}^2=dw d\bar w.}

\ds\ has an infinite number of asymptotic symmetries,
whose associated bulk vector fields $\zeta$
generate the conformal group on \hi \dscft. With
each of these symmetries there is an associated charge.
A general procedure for constructing such charges for spacelike slices
ending on a boundary was given in
\by, adapted to AdS in \bk, and adapted to dS in \dscft.
For \ds\ in planar coordinates, \im\ is a plane and  the charges are
\eqn\dkx{\eqalign{ L_n&={1 \over 2 \pi i}\oint dz\, T_{zz}z^{n+1},\cr
 \bar L_n&=-{1 \over 2 \pi i}\oint d \bar z\,
T_{\bar z\bar  z}\bar z^{n+1},}}
where $T_{zz}$ is the boundary stress tensor given by
\refs{\by,\bk,\dscft}
\eqn\gtyz{T_{\mu \nu}={1 \over 4 G}\bigl[K_{\mu\nu}-
{(K+1)}\gamma_{\mu\nu}\bigr].}
Here $\gamma_{\mu\nu}$ is the induced metric on the boundary, and the
extrinsic curvature is defined by $K_{\mu\nu}=\half {\cal L}_n
\gamma_{\mu\nu}$ with $n^\mu$ the future-directed unit normal.
The  contour integral is over the $S^1$ boundary of  \im\ in planar
coordinates at $|z|=\infty$.
The AD mass \ad\ is proportional to $L_0+\bar L_0$.
The complex coordinates on the boundary cylinder in \rfzl\ are related
to those of the plane by
\eqn\zwi{z=e^{-iw}.}

In the previous section charges ${\cal L}_0$ and  $\bar {\cal L}_0$
were constructed for weak scalar field excitations on a fixed de Sitter
background. These can be related to the weak field limit of
$L_0$ and $\bar L_0$ by using the conservation equation \by\
\eqn\cnsv{{1\over 2\pi}\nabla^\mu T_{\mu\nu}= n^\mu {\cal T}_{\mu\nu},}
which states that the failure of $T_{\mu\nu}$ to be conserved is given by
the matter flux across the boundary. Contracting both sides of
\cnsv\ with a Killing vector $\zeta$ and integrating over a
disc $\Sigma_C$ spanning a contour $C$ on \im\ yields
\eqn\gth{{1\over 2\pi} 
\int_C d\sigma^\mu T_{\mu\nu}\zeta^\nu = \int_{\Sigma_C}
d\Sigma^\mu {\cal T}_{\mu\nu} \zeta^\nu ,}
where $d\sigma^\mu$ is the normal boundary volume element normal to
the curve $C$.  Comparing with \haman, \ccl, and \qzck, we see that
integrand on the right hand side of this expression for $L_0$ ($\bar
L_0$) agrees with that in the expression for ${\cal L}_0$ ($\bar {\cal
L}_0$).
\foot{Our sign convention in \dkx\ was
chosen so that in the weak field limit ${\cal H}$ reduces to the
integral of the scalar stress energy density, without a relative minus
sign.  This convention agrees with \refs{\dscft,\klemm,\lectures}, but
differs by a sign from \refs{\bdm,\ulf}.}  Of course when the fields
are not weak there are gravitational corrections to the bulk
expressions.

The cylinder charges corresponding to \dkx\ are \foot{A minus sign
arises in this expression from the relative orientation of the $z$ and $w$
contours.}
\eqn\dkxm{ H_n=-{1 \over 2 \pi }\int_0^{2 \pi} dw\, T_{ww}e^{-inw}}
and its complex conjugate. We have used the symbol $H_n$ rather than
$L_n$ because on the cylinder \dkxm\ includes a Casimir energy
contribution for $H_0$. We will be interested in $H_0$, which is the
charge associated to the vector field
\eqn\fdt{\zeta_0=\half \left( \p_t+i\p_\varphi\right).}
For $r\to \infty$ one finds
\eqn\tww{T_{ww}={1 \over 4 G}\gamma_{ww}={\tau \over 16 G}.}
Integrating around the cylinder then gives
\eqn\kln{H_0=-{1 \over 2 \pi}\int_0^{2\pi}d\varphi\,T_{ww}=
 -{c \over 24}\tau,}
and similarly
\eqn\kbln{\bar H_0= -{c \over 24}\bar \tau.}
For later convenience
we have written these expressions in terms of the
\ds\ central charge
\eqn\cvl{c={3 \ell \over 2G},}
where we have restored the factor of the de Sitter radius $\ell$.
However, so far our discussions have been purely classical.

We note for pure de Sitter space ($M=1$ and $J=0$) $H_0=-{c \over
24}$. This has a nice interpretation in the dual field theory on the
boundary, as discussed in \klemm.\foot{An alternate interpretation was
given in \bdm.} According to \dscft, the bulk gravity state on the
slice $t=\infty$ in planar coordinates is dual to a CFT state on the
$S^1$ boundary of \im\ (i.e., where the slice $t=\infty$ intersects
\im ) at $z=\infty$. This state is the wave functional produced by fixing
boundary conditions on the $S^1$ and then doing the CFT
path integral over the disc. This should give the $SL(2,C)$ invariant
ground state of the CFT. Transforming from planar to static coordinates
in the bulk is then dual to the conformal mapping from the plane to the
cylinder. This mapping should produce, via the Schwarzian in the 
stress tensor transformation law,  the Casimir energy $-{c \over 24}$
for a CFT with central charge $c$ on a circle of radius $1$.
Indeed this agrees beautifully with the fact that the boundary stress tensor
vanishes in planar coordinates but gives $H_0=-{c \over 24}$ in static
coordinates.

We note for future reference that the state so constructed on \im\ is a
pure state with no entropy.

The agreement with the CFT picture persists for general $\tau$. \kln\
is then precisely the Casimir energy from conformal mapping from the
plane to a cone. We note also that as $M$ decreases, the energy $H_0$
increases, in accord with the expectation that a positive deficit
angle has a positive mass.

\newsec{Entropy}

In this section we discuss the conditions under which the entropy
\sbh\ might be microscopically derived from a 2D CFT. Related
discussions have appeared in \refs{\banados,
\park,\bdm}.

Consider the canonical partition function of a 2D CFT with complex
potential $\beta$,
\eqn\zlls{Z= \int dL_0\, d\bar L_0\,
 \rho(L_0,\bar L_0)\, e^{-\beta L_0-\bar\beta\bar L_0},}
where $\rho$ is the density of states.
We wish to evaluate this in the saddle point approximation.  Let us
assume that we are in a regime where the thermodynamic approximation
is valid, and we can use Cardy's formula \cardy\ for the density of
states\foot{Since we are working in the canonical, rather than
microcanonical picture, the final formula for the entropy is
unaffected by the shift of $L_0$ in the exponent.}
\eqn\dens{\rho(L_0,\bar L_0) = \exp\left[
    2\pi \sqrt{{ c \over 6}(L_0-{c \over 24}) } +
    2\pi \sqrt{{ c \over 6}(\bar L_0-{c \over 24}) } \right],}
When $\beta$ is complex, \zlls\  has a complex saddle point at
$L_0={\pi^2 c \over 6 \beta^2}+{c \over 24}$.\foot{For pure
\ds\ this is $L_0={c \over 12}$, as in
\banados.}
Evaluating the integral at the saddle point and using
$S=(1-\beta  \p_\beta- \bar\beta  \p_{\bar \beta} )\ln Z$ gives
\eqn\dfzl{S=  {\pi^2 c \over 3 \beta} +
{\pi^2 c \over 3 \bar \beta} .}
%
If we now use the formula
\eqn\cvl{c={3 \ell \over 2G},}
for the central charge of the boundary CFT, together with the formula
\eqn\trc{\beta ={2 \pi \over \sqrt{M-i(8GJ)}},}
derived in section 6.3 for the complex temperature of Kerr-\ds , the
microscopic formula \dfzl\ reproduces exactly the macroscopic formula \sbh\
for the Bekenstein-Hawking entropy of Kerr-\ds.

This yields a two-parameter fit relating the area of the Kerr-\ds\
horizon to the number of microstates of a 2D CFT.  However with our
current understanding, this should be regarded as highly suggestive
numerology rather than a derivation of the entropy. One problem is
that the dual CFT is not unitary, and hence is not obligated to obey
Cardy's formula. A second problem is that we have not specified where
the CFT density matrix resides whose entropy is being computed.  In
most discussions---including ours---the quantum state on global de
Sitter is in a pure state. Furthermore its dual---as discussed at the
end of the previous section---is the $SL(2,C)$ invariant CFT vacuum.  A
density matrix arises only after tracing over a correlated but
unobservable sector.  We saw in section 6.3 that for a scalar field in
the (pure) Euclidean vacuum state, a thermal density matrix arises
after a northern trace over the Hilbert space in the unobservable
northern diamond. One might expect that the quantum state of the
boundary CFT would also become thermal after performing a similar
trace.  However it is not clear to us exactly what a northern trace
corresponds to in the boundary CFT on ${\cal I}^\pm$.

It appears that de Sitter entropy arises when attention is restricted
to the true observables in the theory. The boundary CFT includes
information about correlators at acausal separations that do not
directly correspond to observable data.  It is a challenging and
important problem to understand what are the true observables in the
language of the of the boundary CFT.

 \centerline{\bf Acknowledgements} We are grateful to M.~Aganagic,
T.~Banks, M.~Headrick, G.~Horowitz, A.~Karch, G.~Moore, M.~Spradlin,
N.~Toumbas, A.~Volovich and E.~Witten for useful conversations. This
work was supported in part by DOE grant DE-FG02-91ER40654, an NSF
Graduate Fellowship, and by the National Science Foundation under
Grant No.\ PHY99-07949.

\appendix{A}{Alternate Forms of Green Functions on $dS_{d}$}

In this Appendix we present several alternate expressions for the
Green functions.

First, let us consider a de Sitter invariant vacuum $|\Omega\rangle$,
so that the wave equation for $G_\Omega(x,x')$ becomes
\eqn\geq{ (1-P^2) \p^2_P G - dP \p_PG -m^2 G = 0,}
where $P$ is related to the geodesic distance $\theta(x,x')$ by
\eqn\thet{ P = \cos \theta.}
Note that if $G_{d,m^2}$ solves \geq\ in $d$ dimensions for mass-squared
$m^2$, then $\p_PG_{d,m^2}$ solves \geq\ in $d+2$ dimensions with
mass-squared $m^2+d$.  This gives an iterative procedure for constructing
Green's functions in all dimensions.  We find
\eqn\iter{\eqalign{
G_{3+2n,m^2} &= \p_P^n G_{3, m^2+1-(n+1)^2} \cr
G_{2+2n,m^2} &= \p_P^n G_{2, m^2-n(n+1) }
}}
where $n$ is a positive integer.

Let us first consider odd $d$.  For $d=3$,
if we let
\eqn\chidef{ G_{3,m^2} = {\chi \over \sin \theta} }
then $\chi$ satisfies
\eqn\chieq{ \p^2_\theta \chi +(1-m^2) \chi = 0.}
So the general solution in $3$ dimensions is
\eqn\gis{ G_{3,m^2} =
{A\sinh\mu (\pi-\theta) + B \sinh\mu \theta \over \sin\theta}}
where $\mu=\sqrt{m^2-1}$ and $A$ and $B$ are arbitrary constants.
The first term gives the usual short distance singularity for the Euclidean
vacuum---with the correct normalization, it gives the usual expression
\gcd.  The second term is present for the transformed vacuum states $\av$,
and has the antipodal singularities mentioned in section 2.2. From
\iter\ we can obtain an expression for the Green functions in higher
dimensions,
\eqn\godd{G_{d,m^2} =
    \sum_{m=0}^n {n \choose m} { \Gamma(n-m+2i\mu)\over \Gamma(m+1+2i\mu) }
{\left(A\sinh(2\mu-in+2im)(\pi-\theta) + B\sinh(2\mu-in+2im)\theta\right)
\over \sin^{d-2}\theta}
}
where $n=\half (d-3)$ and 
\eqn\muis{\mu=\sqrt{m^2-\left({d-1\over2}\right)^2}.}
We have absorbed an overall normalization into the constants $A$ and $B$.
As a function of $\theta$, $G$ has
isolated singularities but no branch cuts.  However, $\theta = \cos^{-1} P$
has a branch cut from $P=1$ to $\infty$ along the real axis, across which
$\theta(P)$ changes sign.
When expressed as a function of $P$, $G$ will likewise have a branch cut.

For even $d$, we start with the $d=2$ solution in terms of Legendre
functions
\eqn\gtwo{
G_{2,m^2} = A P_\nu (\cos\theta) + B Q_\nu (\cos\theta)
}
where $\nu(\nu+1) = -m^2$.
So
\eqn\geven{G_{d,m^2} = 
A P^{(n)}_\nu (\cos\theta) + B Q^{(n)}_\nu (\cos\theta)}
where $n=\half(d-2)$ and $\nu(\nu+1)=n(n+1)-m^2$.  Here, $P^{(n)}_\nu$
is an associated Legendre function,
the $n^{th}$ derivative of the Legendre function.

\appendix{B}{Properties of Hypergeometric Functions}

We collect a few relevant facts about
hypergeometric functions.  More details may be found in, e.g.,
\abramowitz.

The formula
\eqn\trfm{F(a,b;c;z)=(1-z)^{c-a-b}F(c-a,c-b;c;z)}
relates hypergeometric functions of $z$ with different values of
parameters, as in \gji.
To relate hypergeometric functions of different variables we use
\eqn\svxx{\eqalign{
F(a,b;c;z)&=
{\Gamma(c) \Gamma(b-a) \over  \Gamma(c-a) \Gamma(b)} (-z)^{-a}
F(a,a+1-c;a+1-b;{1\over z})\cr
&~~~~~~~~~~+{\Gamma(c) \Gamma(a-b) \over  \Gamma(a) \Gamma(c-b)}
 (-z)^{-b}F(b,b+1-c;b+1-a;{1\over z})\cr
&={\Gamma(c) \Gamma
(c-a-b) \over  \Gamma(c-a) \Gamma(c-b)}
F(a,b;1+a+b-c;{1 - z})\cr
&~~~~~~~~~~+{\Gamma(c) \Gamma(a+b-c) \over  \Gamma(a) \Gamma(b)}
(1-z)^{c-a-b}F(c-a,c-b;c-a-b+1;{1 - z}).}
}
These give us the Bogolyubov relations \bdef\ and \rel, respectively.
Since $F(a,b;c;0)=1$
these equations also fix the behavior of $F(a,b;c;z)$ as $z\to\infty$
and $z\to1$, as in \tzsf, \fhxddr\ and \fhxr.

\listrefs

\end